# The Radio-wave Observations at the Lunar Surface of the photoElectron Sheath (ROLSES) instrument onboard the Intuitive Machines-1 Mission to the Moon

Nat Gopalswamy[1], Robert J. MacDowall[1,*] Pertti Mäkelä[1], Scott A. Boardsen[1], Seiji Yashiro[1], Richard B. Katz[1], Igor Kleyner[1], Scott D. Murphy[1], Richard C. Mills[1], Chimaobi Onyeachu[1], Michael K. Choi[1], Thomas M. Schluszas[1], Victor Gonzalez-leon[1], Pietro A. Sparacino[1], William M. Farrell[1,2,3], Damon C. Bradley[1,3], Jack O. Burns[4], Joshua J. Hibbard[4,5]

[1]*Heliophysics, NASA/GSFC, 8800 Greenbelt Rd, Greenbelt MD 20771, USA, nat.gopalswamy@nasa.gov*
[2]*Currently at Space Science Institute, 4765 Walnut St, Suite B, Boulder, CO 80301, USA*
[3]*Currently at DeepSpace Technologies. 8865 Stanford Blvd, Suite 183. Columbia, MD 21045, USA*
[4]*Department of Astrophysical and Planetary Sciences, University of Colorado, Boulder, CO 80309, USA*
[5]*Currently at Space Sciences Lab, University of California, Berkeley; Berkeley, CA 94720, USA*

The Radio-wave Observations at the Lunar Surface of the photoElectron Sheath (ROLSES) instrument is a radio telescope system designed to characterize the radio and plasma wave environment of the near-side lunar surface at frequencies between 2 kHz and 30 MHz. The ROLSES sensor consists of a set of four 2.5 meter radio monopole antennas onboard the Intuitive Machines – (IM-1) lander, Odysseus. The antennas were stowed during launch to be deployed after landing on the lunar surface using a frangibolt mechanism. The frequency range is well suited to observing radio waves at frequencies below 15 MHz that cannot be observed from Earth due to the ionospheric cutoff. The higher frequencies are helpful in cross-calibrating with ground-based observations. Radio waves from the Sun, the Milky Way galaxy, Jupiter, Earth's auroral region, and ground-based radio transmitters were expected to be present on the lunar surface. In addition, local plasma fluctuations and lunar/interplanetary dust were expected to produce detectable radio signals. Radio data from each of the 4 antennas, after passing through an isolating pre-amp and signal conditioning analog electronics, are digitized to 14 bits at 120 mega samples per second and then digitally processed by a Field Programmable Gate Array (FPGA) that performs onboard spectral analysis via a Fast Fourier Transform (FFT). Time-averaged spectral values are then stored and returned. ROLSES science data are sent to the lander at ~ 16 kb/s and subsequently downlinked for further processing. Also telemetered to Earth are raw waveforms (unprocessed time sequence data) that are useful in studying dust impact on ROLSES antennas. ROLSES data are sent to the lander and subsequently downlinked for further processing. ROLSES is part of NASA's Commercial Lunar Payload Services (CLPS) program. Odysseus landed close to the south pole at Malapert A (~80S). This paper describes the complete design and operations of the ROLSES instrument and presents initial observations. Despite the tilted landing of Odysseus, we were able to successfully command, deploy the antennas, and operate the instrument in transit and on the lunar surface. We also describe an enhanced version of this instrument (ROLSES 2) currently under development.

## 1. Introduction

At radio frequencies below ~30 MHz, a rich variety of radio waves can be observed from space. Nonthermal radio emissions at these frequencies tell us about acceleration of electrons in various astrophysical situations. For example, radio bursts from the Sun provide key insights into electron acceleration in current sheets and in shock fronts. By observing these radio bursts, we learn about the acceleration processes and the properties of the ambient corona and interplanetary medium through which solar disturbances propagate. Similarly, energetic electrons accelerated in our Galaxy produce synchrotron emission observed as a background continuum. Accelerated electrons in Earth's auroral regions produce radio emission at kilometer wavelengths via the electron cyclotron maser mechanism. Radio emission from Jupiter informs us about electron acceleration in the Jovian atmosphere. All these radio emissions are remote sensed by space radio instruments because of the ionospheric cutoff.

Recognizing the importance of the ionospheric plasma, Canadian satellites Alouette-I and II were launched in the beginning of 1960s to map the ionosphere in the frequency range 1-10 MHz (Hartz, 1964a,b). These were followed by the International Satellites for Ionospheric Studies (ISIS) program that launched two more satellites with an expanded frequency range (0.1-20 MHz). Although the primary mission of the Alouette and ISIS radio instruments was to map the ionospheric density, they detected solar type III radio bursts and galactic background from which the spectra of these phenomena could be derived. While the low-frequency observations from low earth orbit was the key aspects of these early missions, ideas on lunar-based, low-frequency radio astronomy also started emerging. Gorgolewski (1966) proposed a lunar surface-orbiter interferometer that would image cosmic radio sources, free from the ionospheric effects and terrestrial transmitter signals. A few years later, the Radio Astronomy Explorer -2 (RAE2) was placed in lunar orbit and contributed to galactic, solar, and planetary astronomy (Alexander et al. 1975). Over the next two decades, many concepts for low frequency radio observatories from the Moon were advanced (see a review by Burns et al. 2019; Chen et al. 2019).

Among the post-Apollo lunar missions, the Kaguya mission from Japan stands out because of the measurement of electron density in the lunar ionosphere using its Radio Science experiment (Imamura et al. 2012). This was basically a radio occultation experiment at much higher frequencies (S and X band). Kaguya found an electron density of 100-300 $cm^{-3}$ on the sunlit side. This was also confirmed by Chandrayaan-1 using radio occultation technic, indicating an electron density of ~300 $cm^{-3}$ close to the lunar surface (Choudhary et al. 2016). Chandrayaan-2 found surprisingly high electron densities on the nightside probably due to plasma trapped in magnetic anomaly regions (Tripathy et al. 2022).

Landed payload missions carrying low-frequency radio instruments have begun in the past decade starting with the Chang'E-4 mission to the farside of the Moon (Chen et al. 2019). Launched in late 2018 and landing in January 2019, Chang'E-4 carried a Low-Frequency Radio Spectrometer (LFRS) on the lander and the Netherlands-China Low Frequency Explorer (NCLE) on the relay satellite Queqiao for radio astronomical observations. The mission demonstrated that the Moon's farside is radio quiet with only RFI coming from the lander itself. Chang'E-4 mission also performed a farside SETI (search for extra-terrestrial intelligence) without Earth's RFI (Li et al. 2026). NASA's Commercial Lunar Payload Services (CLPS) program, which is a high risk/high reward program that provides regular access to the lunar surface 2-3 times per year. The Radio-wave Observations at the Lunar Surface of the photoElectron Sheath (ROLSES) instrument (described in this paper) represents a pioneering attempt to characterize the radio-radiation environment of the Moon. ROLSES was the first NASA radio science payload operating in the frequency range 2 kHz to 30 MHz. deployed under the CLPS program. ROLSES was designed to provide useful information on the development and observational strategy for other radio instruments such as the LuSEE-Night, which will make the first nighttime observations of the radio band (0.1-50 MHz) corresponding to the early Universe's Cosmic Dark Ages (Bale et al. 2023). LuSEE-Night, developed as a joint collaboration between NASA and the U.S. Department of Energy, is scheduled to land on the farside of the Moon in 2027. A lunar orbiter experiment known as PRATUSH (Probing ReionizATion of the Universe using Signal from Hydrogen) is under conceptual development to detect the redshifted global 21-cm signal from the Cosmic Dawn and Epoch of Reionization in the frequency range 55 - 110 MHz (Sathyanarayana Rao et al. 2023). Although signals at these frequencies are accessible from ground, the mission would exploit the radio-quiet environs of the lunar farside. Another concept being developed is the Space Electric and Magnetic Sensors (SEAMS) mission, which will operate in the frequency range 0.3 to 16 MHz in LEO with the ultimate goal of placing the sensors on a lunar orbiter (Tanti et al. 2023).

In this paper, we describe the instrument and present initial results obtained from the observations made during the 2024 Intuitive Machines - 1 (IM-1) mission that carried ROLSES. ROLSES is one of the NASA Provided Lunar Payloads (NPLPs) selected under the CLPS program. The CLPS program is designed to bring down the cost for science investigations and technology demonstrations going to the Moon and to make them more routine in the lead-up to the Artemis landings. ROLSES was one of the six NASA instruments manifested on the IM-1 lander but the only one dedicated to collecting science data. IM-1 was launched using a SpaceX Falcon 9 rocket from Launch Complex 39A at the Kennedy Space Center on February 15, 2024 at 06:05:00 UT and landed near the Lunar south pole on February 22, 2024 at 23:23:53 UT - the first U.S. lunar landing in more than 50 years. The target landing location was Malapert A (-80.20594, 0.93919) on the near side ~300 km from Moon's south pole. The lander ended up at a location, about 1.5 km from the target site and was tipped over with a broken leg . During the final trajectory maneuver enroute to Moon, one of the antennas received excessive solar flux that overheated the frangibolt resulting in the inadvertent deployment and damage of the antenna when the lander was about 9600 km from the Moon. Upon landing, another antenna deployed without commanding, again owing to excess solar flux due to the tilted lander. Despite the lander troubles and inadvertent deployments, it was possible to deploy the remaining two antennas and operate ROLSES to collect some useful data. Detailed data analysis and first science results were recently reported in Hibbard et al. (2026).

Section 2 summarizes the scientific goal and objectives of ROLSES. Section 3 describes the ROLSES instrument with details on each of the major elements. Section 4 describes launch and landing of the IM-1 mission including scientific operations. Section 5 describes ROLSES data processing, while section 6 presents the initial results. Section 7 describes the changes and improvements made in the next version of the instrument (ROLSES 2). Finally, Section 8 summarizes the paper.

## 2. Goal and Objectives

The primary science goal of ROLSES is to understand the radio-radiation environment of the lunar surface for further exploration and development of radio observatories on the Moon. To advance this goal, the following objectives have been set.

(1) Detect and characterize solar and planetary radio waves from a lunar surface observatory. The solar radio bursts are early indicators of solar transients, some of which can adversely affect Earth's space environment.

(2) Detect terrestrial natural auroral and anthropogenic radio emissions, to thus assess the Earth as a 'noisy' radio source.

(3) Measure the electron density in the vicinity of the lander via the local electron plasma frequency oscillations.

(4) Sense interplanetary and 'slow moving' lunar dust via grain contacts with the antenna.

(5) Search for evidence of detection of galactic background radiation.

(6) Assess the radio frequency interference (RFI) from the lunar lander, to thus determine the suitability of the lander as a radio observation platform.

In the following we provide additional information on the objectives, achieving which will result in an improved understanding of the low-frequency universe.

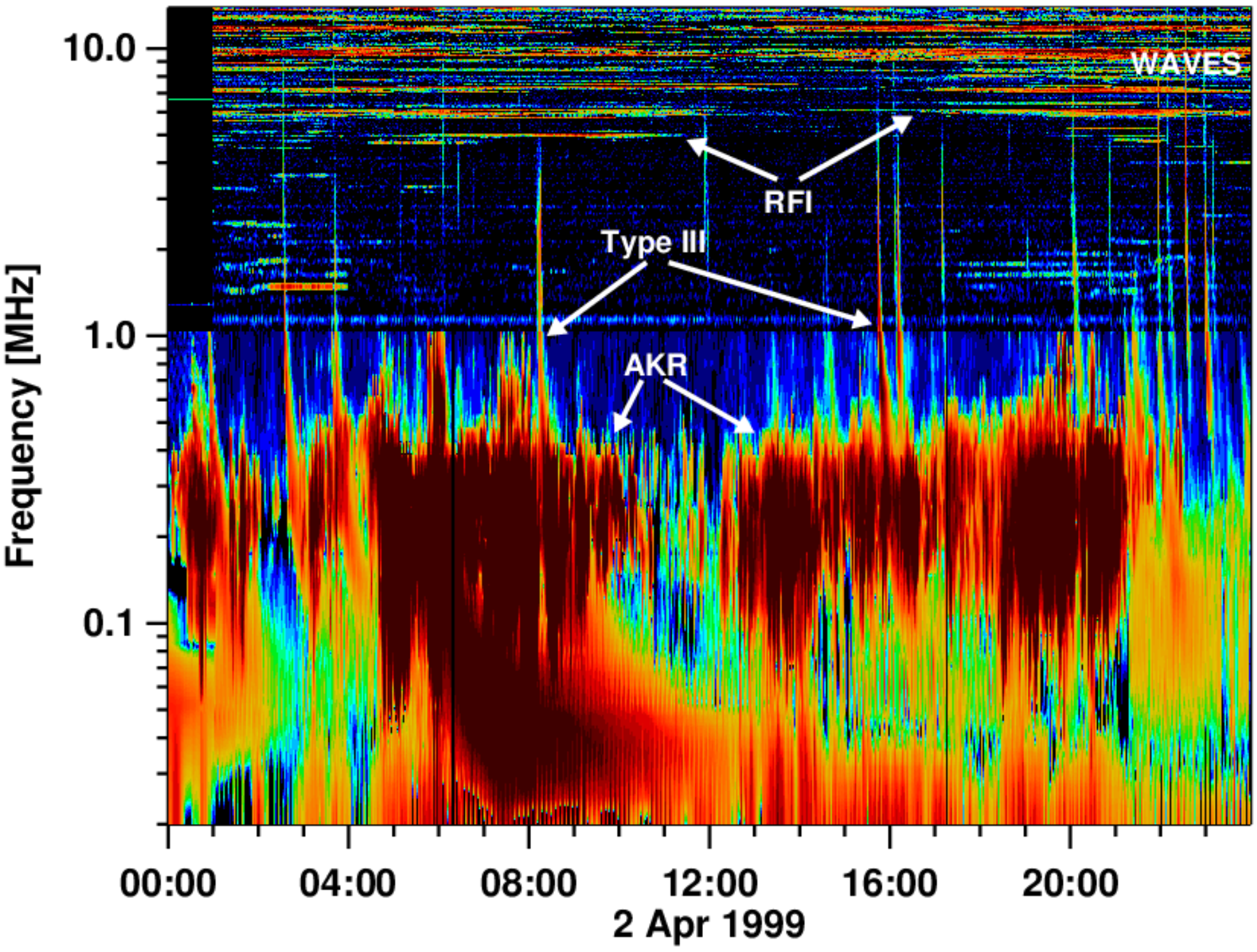

Fig. 1. Wind/WAVES dynamic spectrum obtained on April 2, 1999 when the Wind spacecraft was in the vicinity of the Moon. Radio frequency interference (RFI), solar type III radio bursts, and the auroral kilometric radiation (AKR) are marked.

### 2.1 Solar Radio Bursts

While radio probes in the interplanetary space have been making measurements at frequencies below 2 MHz since the 1960s, the radio and plasma wave (WAVES) experiment on board the Wind (Bougeret et al. 1995), STEREO (Bougeret et al. 2008), Parker Solar Probe (Bale et al. 2016), and Solar Orbiter (Maksimovic et al. 2020) have the additional capability to observe phenomena in the frequency range 2-15 MHz. The Moon provides another platform to observe the low frequency radio waves. Electromagnetic fluctuations in the vicinity of antennas on the Moon and dust impacts on the antennas also result in detectable radio signals, which can be studied to obtain information on the physical processes in the local environment of the Moon.

At the Sun, the low frequency range corresponds to plasma frequencies in the outer corona and hence probe solar disturbances just leaving the Sun that likely impact planets and spacecraft in the heliosphere. There are four types of solar radio bursts that dominate in the ROLSES frequency range: Type II, III and IV bursts from solar eruptions, and type III storms from active regions outside of solar eruptions (Gopalswamy 2011). The type III bursts seen in Fig. 1 are due to isolated energy releases on the Sun (small flares) resulting in the acceleration of electrons that produce the observed radio emission via the plasma emission mechanism. Type III bursts can occur as a group and last for tens of minutes during large-scale solar eruptions involving coronal mass ejections (CMEs) and major flares. Fast CMEs drive fast mode MHD shocks that accelerate electrons to produce type II bursts. While the type III bursts are due to electrons from the flare site propagating along open magnetic field lines, type IV bursts are produced when some of the accelerated electrons get trapped in closed field lines. The ROLSES frequency range extends to higher frequencies for comparison with ground-based observations. Figure 2 is another Wind/WAVES dynamic spectrum showing a pair of eruptions with all the four types of solar bursts. Phenomena unique to the low-frequency regime are the start of some interplanetary type II bursts, end of type IV bursts, and disruption of type III storm by CMEs. The eruption-associated bursts are key indicators of large-scale CMEs leaving the Sun that can cause large solar energetic particles and geomagnetic storms. These phenomena are important space weather events that can affect humans and their technology in space, including those on the surface of the Moon.

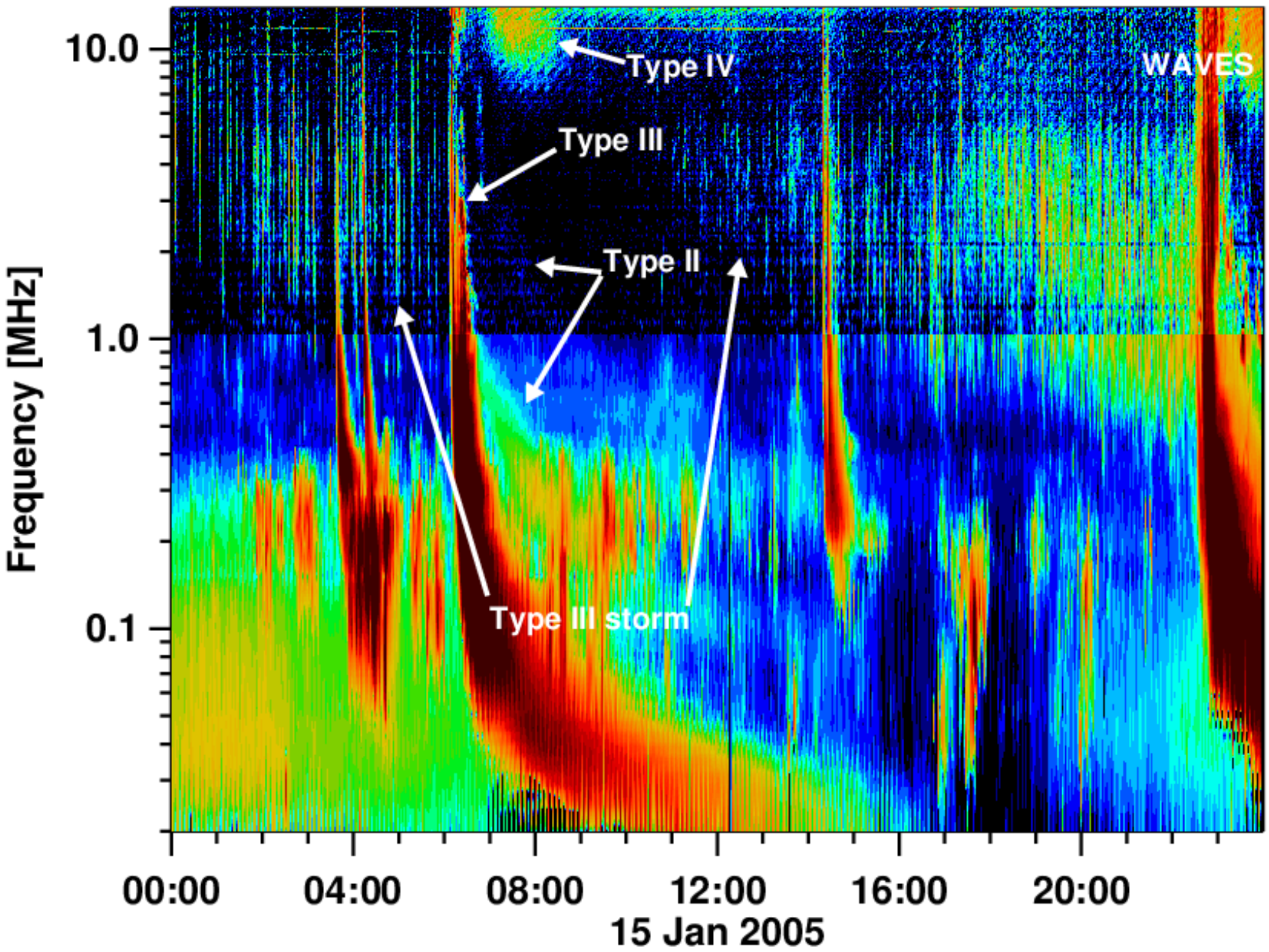

Fig. 2. Wind/WAVES dynamic spectrum when Wind was at Sun-Earth Lagrange point L1. No terrestrial RFI is detected because the spacecraft is located at a large distance from Earth (~$1.5{\times}10^6$ km). Eruption bursts of type III, Type III, and type IV are marked. Also marked are type III storms, which consist of a large number of short-duration bursts occurring quasi continuously. The individual bursts resemble regular type III bursts, but much shorter in duration and weaker in intensity. The type III storm is disrupted by the eruption around 6:30 UT followed by its reappearance around 12:00 UT. It was further disrupted by a second eruption towards the end of the day.

## 2.2 Planetary Radio Emission

Among planetary radio emissions, the Jovian decametric emission (DAM) is of particular interest because of the noteworthy mechanisms of particle acceleration and radio wave generation. DAM is typically observed in the range 3 MHz to 40 MHz, with a spectral peak around 10 MHz (Clarke et al. 2014). The DAM emission frequencies thus have a good overlap with the ROLSES frequency range of 2 kHz to 30 MHz. The auroral kilometric radiation (AKR) is another type of planetary radio emission thought to be produced by the interaction between Earth's magnetic field and accelerated electrons in the Earth's polar region. AKR is indicative of the solar wind conditions (speed, magnetic field, geomagnetic activity). Based on a study of ~5000 AKR bursts, it was found the bursts have a median lifetime of ~30 min (Fogg et al. 2022). AKR has never been measured from the lunar surface. This unique perspective will offer new insights into the nature and physics of the cyclotron maser instabilities that are thought to result in AKR. These observations will provide models for similar emissions from the magnetospheres of exoplanets that can be measured by a lunar far side interferometer.

In addition to AKR, which is a natural emission from the vicinity of Earth, one expects to observe terrestrial anthropogenic radio interference at the Moon from ground-based radio transmitters. In fact, when the Wind spacecraft was in the vicinity of the Moon on April 2, 1999, clear RFI from ground transmitters were detected by the Radio and Plasma Wave experiment (WAVES). Figure 1 shows a frequency vs time radio spectrogram from the Wind/Waves radio instrument showing intense terrestrial radio transmitter RFI at frequencies above 6 MHz (labeled as 'RFI'). There are brief periods when RFI can be seen at lower frequencies caused by density irregularities in the ionosphere. Also detected by Wind during this time are Type III radio bursts appearing as broadband (vertically-appearing) short-duration events that originate from the Sun (i.e., not affected by the terrestrial ionospheric cutoff). Also labeled in Fig. 1 is intense AKR detected between 200-800 kHz that originates from the Earth's auroral regions at altitudes above the polar ionosphere. Despite the RFI, solar bursts can be readily distinguished due to their broadband bursty morphology in the dynamic spectrum. While Wind/WAVES is on a free-flying spacecraft, ROLSES was on a fixed platform, the Moon. The lander operations can also cause local RFI, which needs to be characterized as well to fully understand the radio environment on the Moon. It is evident from Fig. 1 that there are a number of dynamic sources of radio emission detectable at lunar distances. Future landed lunar radio observatories searching for weak cosmic sources will have to contend with this complex radio environment.

### 2.3 Electron Density near the Lunar Surface

The exo-ionosphere of the Moon is also poorly characterized. Photoelectrons are produced on the lunar surface, and their density falls off rapidly in the vertical direction. The fluctuations in the electron density peak around the local plasma frequency in the vicinity of the antenna and hence can be sensed by an antenna. Measuring these electric field fluctuations at the local electron plasma frequency provides a measure of the electron density and is important to determine the effect on the antenna response of larger lunar radio observatories. On a frequency vs time spectrogram, these electric field fluctuations are intrinsically narrow banded, with a center frequency very close to the electron plasma frequency. Consequently, the emission provides a highly resolved measurement of the local electron density, since the electron density varies as the square of the electron plasma frequency. The intensity of this narrowbanded emission in the local photoelectron sheath is not known, but if the sheath itself is dynamic then the local plasma oscillations may become intense. Such a narrowbanded, time-varying should be apparent in frequency vs time spectrograms, forming horizontal 'wiggly lines' on a spectrogram. Such an emission stands out relative to broadbanded signals. Simulations indicate that the photoelectron density and its vertical profile vary significantly as a function of height above the lunar surface for various solar wind environments such as flares, CMEs, and ordinary solar wind (Farrell et al. 2013). At a height of ~1 m above the surface the electron density has been inferred to be ~ $5 \times 10^7$ $m^{-3}$ (electron plasma frequency ~ 64 kHz). Following a shock passage, the density falls by a factor of ~2 but increases by a factor of ~ 2 during a CME. These electron densities translate to a plasma frequency range of 10-300 kHz, which corresponds to the lower frequency band of ROLSES. A set of these simulations also indicated that the sheath itself can be very dynamic with the escape of photoelectrons occurring in periodic releases associated with changes in the local surface potential controlled by the periodic return of trapped electrons. Such dynamic activity is expected to trigger intense local plasma oscillations.

### 2.4 Lunar Dust

When a charged dust grain encounters an antenna, an impulsive charge exchange occurs between the antenna and the grain. The grain is usually at a different potential than the antenna and hence a voltage signal is generated during the impact. Such a technique has been used successfully in detecting triboelectrically charged grains in terrestrial dust devils (Houser et al., 2003; Farrell et al. 2004). The horizon glow observed by Surveyor landers is thought to be dust-scattered light during terminator crossings (see Colwell et al. 2007). ROLSES provides an opportunity to detect the lofted/levitated dust from direct observations. Continuing the ROLSES observations into the lunar night provides the best opportunity for dust detection because the dust impact events are thought to peak near the terminators. In waveform data, the dust-antenna charge transfer appears as short-lived impulse or 'spike' that exceeds the background noise floor. Upon spectral analysis, the waveform spike appears as an isolated broadband event on a frequency-vs-time spectrogram.

## 2.5 Galactic Radio Emission

Going beyond the solar system, the Milky way is an important source of radio emission whose spectrum peaks in the ROLSES frequency range (Hartz 1969; Novaco and Brown 1978). Most recently, the Galactic spectrum was measured between 1 and 6 MHz by the FIELDS instrument on the Parker Solar Probe (Bassett et al. 2023); ROLSES extended the measurements over a larger frequency range and made the first detection of the Galaxy from the Moon (Hibbard et al. 2026). Such measurements will help characterize the foreground emission for 21-cm hydrogen cosmology measurements of the Dark Ages global spectrum (see e.g., Satyanarayana Rao et al. 2023). The galactic background is produced primarily by highly relativistic cosmic rays spiraling in the galactic magnetic field and producing smooth, broad-band synchrotron emission across the entire observing band of ROLSES. As the distribution of particles follows a power-law, the synchrotron radiation exhibits a spectral slope of approximately -2.5, peaking near 1 MHz at around 10^6 Kelvin. At that turnover point, free-free absorption due to electrons in the plane of the galaxy begins to dominate, flattening the slope of the spectrum. The galactic foreground is persistent but highly anisotropic, producing a variable signature in low-frequency radio observations that depends strongly on galactic hour angle and that peaks when the galactic center is above the horizon (for broad-beam antennas). The galaxy in particular would induce a periodic modulation in the data consistent with the galactic hour angle for the observation location. The spectral structure has been well-measured and could also be discerned by its slope, at least for the first several principal components of a spectral data cube.

## 2.6 Lander Noise

One obvious source of RFI is the lander itself and the payloads manifested on it. Landers produce via mechanisms like thruster operations, motors, and digital electronics. These RFIs will also be recorded in the radio spectrum, Data analysis needs to take care of these RFI while interpreting spectra in identifying cosmic sources. The lander RFI is likely to be different for different landers and the collection of payloads with varying operational requirements. For ROLSES, we do expect RFI from the instrument because of the digital signal processing in the spectrometer.

Table 1 outlines the measurement plan for ROLSES to achieve these objectives. Because of issues with the IM-1 lander, only objectives 2, 5, and 6 were actually addressed with ROLSES at its South Pole landing site.

Table 1. Measurement objectives of ROLSES

| Objective/measurement | Frequency range | Frequency resolution | Sampling rate |
|---|---|---|---|
| Detect solar and planetary radio bursts and galactic radio emission (objectives 1, 5) | 10 kHz–30 MHz | 512 samples per Low and High band | 8 sec |
| Detect terrestrial natural auroral and human-made radio emissions (objective 2) | 10 kHz–30 MHz | 512 samples per Low and High band | 8 sec |
| Photoelectron sheath density and evolution (objective 3) | 10-300 kHz | ~ 2 kHz | 8 sec |
| Detect Interplanetary and lunar dust (objective 4) | 10-600 kHz | 512 samples per Low and High band | 8 sec |
| Detect RFI from the lunar lander (objective 6) | 10-KHz-30 MHz | 512 samples per Low and High band | 8 sec |

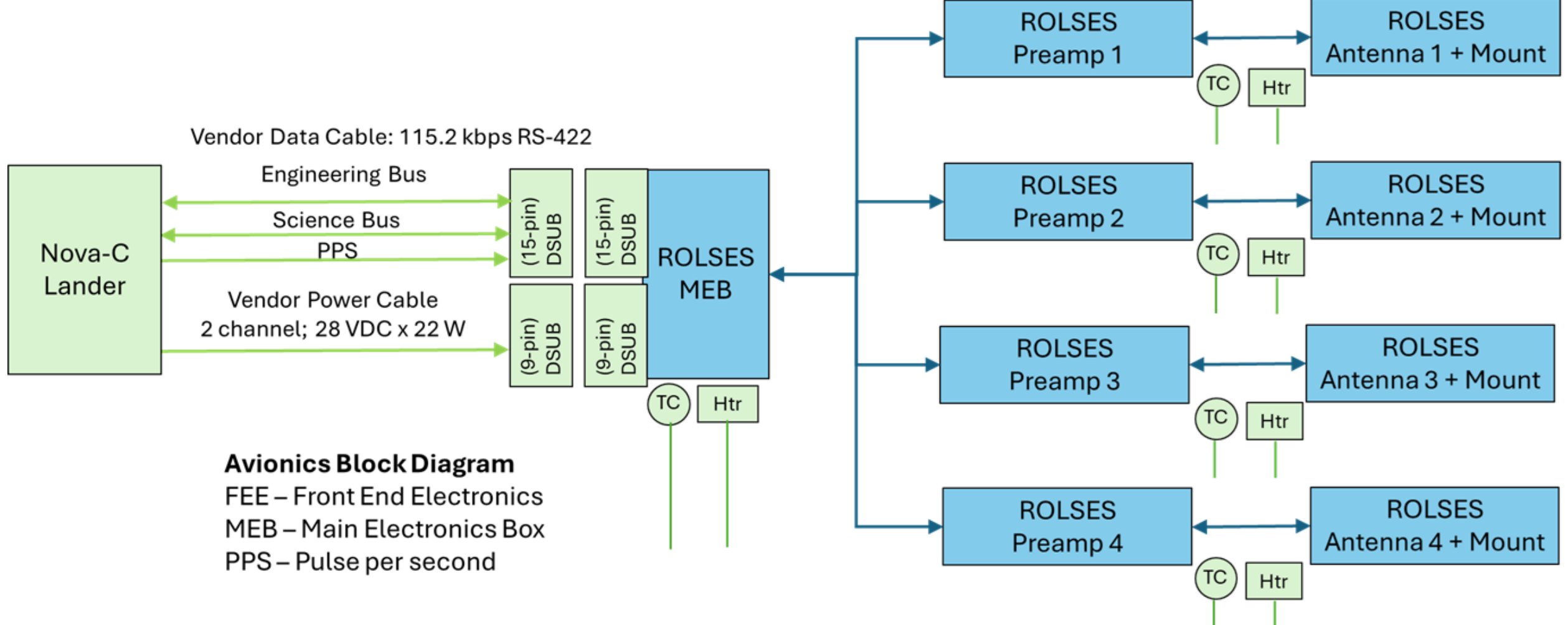

Fig. 3. ROLSES avionics block diagram showing the primary subsystems and concept of operations. The blue sections are the responsibility of the ROLSES team, and the green sections are the responsibility of Intuitive Machines. The command, data, and timing interface to the spacecraft (Nova-C lander, renamed Odysseus) is via RS-422 connections. Commands are sent from the lander to the MEB over the Engineering bus. Power is received from the lander on two 28 VDC feeds, one for powering ROLSES and the other for the sole purpose of deploying antennas. The signals from the four ROLSES monopole antennas are conditioned by the preamps and then sent to the MEB. The MEB packetizes raw data as well as data output from the four digital signal processor (DSP) modules, which performs Fast Fourier Transforms (FFTs). The science and housekeeping data are sent to the lander over the science bus at a rate of 16 kb/s. The lander also provides thermocouples (TC) and 5 W heaters (Htr) to the FEEs and MEB.

## 3. Instrument Overview

ROLSES is a low-frequency radio telescope designed to make unique radio and plasma wave measurements from the lunar surface. The ROLSES payload consists of a radio receiver, fed by a set of four STACER antennas. The antennas sense the radio radiation in the frequency range 2 kHz to 30 MHz. The signals from each of the 4 monopole antennas drive an isolating pre-amp (front end electronics, FEE) and signal conditioning analog electronics. The broadband signals from each antenna are then digitized to 14 bits at 120 mega samples per second. The 4 digitized data streams are then digitally processed by a Field Programmable Gate Array (FPGA) that performs onboard spectral analysis via a Fast Fourier Transform (FFT) in the main electronics box (MEB). The FPGA used was Microchip's RTG4-1. Another choice was SmartFusion2 device, which would have worked fine as well. RTG4-1 is a space-qualified, radiation-tolerant FPGA; SmartFusion2 is a commercial/industrial System-on-Chip FPGA that pairs standard FPGA logic with a hardened ARM Cortex-M3 processor. Anticipating the harsh radiation environment of the Moon, we decided to go with RTG4-1. Time-averaged spectral values from the transforms are then stored in the lander and later returned to Earth. Electric field waveforms from each antenna were also captured and returned. This waveform mode is especially valuable to verify the signals going to the FFT processing system and for dust detection. The FEE outputs are connected to the Main Electronics Box (MEB). The MEB receives deployment and operational power from the lander and sends science and housekeeping data to the lander at 16 kb/s. The ROLSES MEB communicated with the lander using two data channels, one for science data and the other for engineering commands (see Fig. 3). The data were downlinked from the lander to the ground, which was processed by Ground Software and distributed to the science team.

### *3.1 Antennas*

ROLSES uses four monopole STACER (Spring-Tensioned Antenna Construction, Extension, and Retraction) antennas designed and built by Heliospace Inc. The antennas are stowed in cans, attached to the outer surface of the lander, and deployed using a frangibolt mechanism after landing on the Moon. There are two upper antennas and two lower antennas located at ~2.5 m and ~1m above the lunar surface, respectively. Each pair of antennas are horizontally opposed.

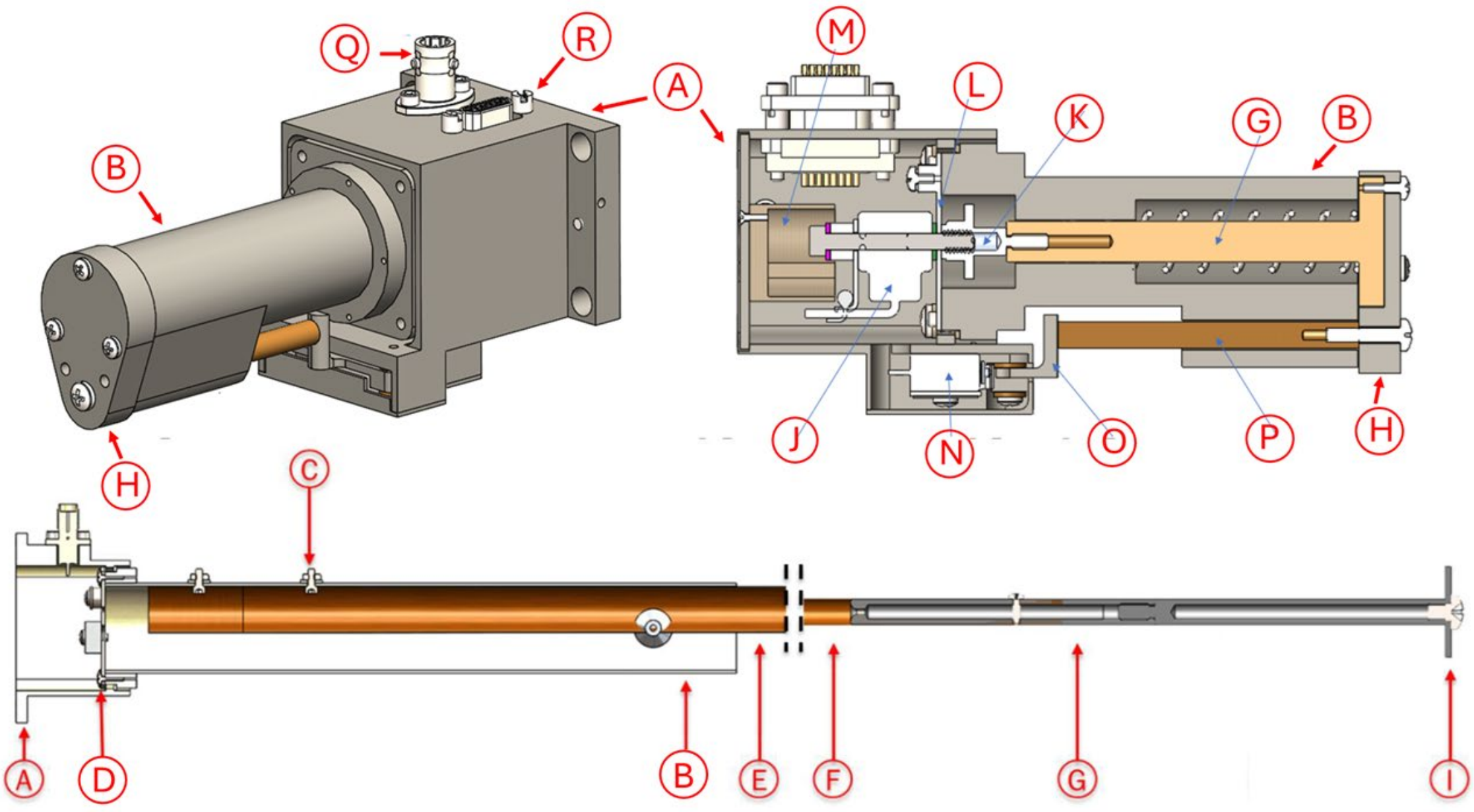

Fig. 4. An overview of the STACER antenna and housing including the frangibolt mechanism. A – Connector housing, B – Stacer can, C – Stacer Fastener, D – Roller fastener, E – Stacer base, F – Stacer, G – Inner tip piece, H – Tip piece cap, I – Outer tip piece, J – Frangibolt, K – Frangibolt stud, L – Can cup, M – Frangibolt cup, N – Micro Switch, O – Switch lever, P – Switch relay plunger, Q – RF 803 Bayonet connector, R – MTD (Modular Twisted-Pair Data) connector.

A STACER, originally developed by Hunter Spring (AMETEK) engineers, is a flat, prestressed metal strip wound into a compact coil. STACER antennas, designed and built by Heliospace Inc., have been used regularly in space applications as booms on which sensors are mounted, and directly as antennas. The STACER used in ROLSES instrument is made of beryllium copper (Be–Cu) alloy, similar to STEREO/WAVES antennas (Bale et al. 2008) but of smaller dimensions. Each antenna has four major components: the STACER, STACER can, the tip piece, and connector/frangibolt housing. The tip piece and housing are made of aluminum. The frangibolt is a non-explosive release device that makes use of a shape memory alloy (SMA) to forcefully break a bolt in tension. TiNi Mini frangibolts are used as actuators for deployment of the antennas. The connector housing provides launch storage, deployment mechanism, and mechanical support at the proper orientation.

The deployed size of the STACER antennas is ~ 2.5 m with a tip diameter of 7 mm and a base diameter of 11.6 mm, giving a surface area of about 0.15 m$^2$. The mass of each STACER element is ~ 0.23 kg (see Table 1). The pre-deployment upper temperature requirement was 70 °C. Hardware and materials associated with the antenna were selected for at least 200° C post-deployment survival. A safety bar was attached to the washer cover to prevent accidental deployment of the STACER; the bars are removed before flight. The frangibolt is positioned inside the connector housing as shown in Fig. 4. The Power and RF connectors were mounted on the outer wall of the housing. The tip piece, STACER, can, and the can mounting flange are collectively referred to as “the antenna” and are electrically isolated from the connector housing.

Table 2. High level specifications of the antennas

| Mass | 0.23 kg |
|---|---|
| Deployed length | 2500 mm |
| Diameter at the base | 11.6 mm |
| Diameter at the tip | 7.1 mm |
| Stowed length | 214.9 mm |
| Acuator | Frangibolt |
| Deployment time | ~1 s |
| Tip deflection (0.17g) | 37.0 mm |

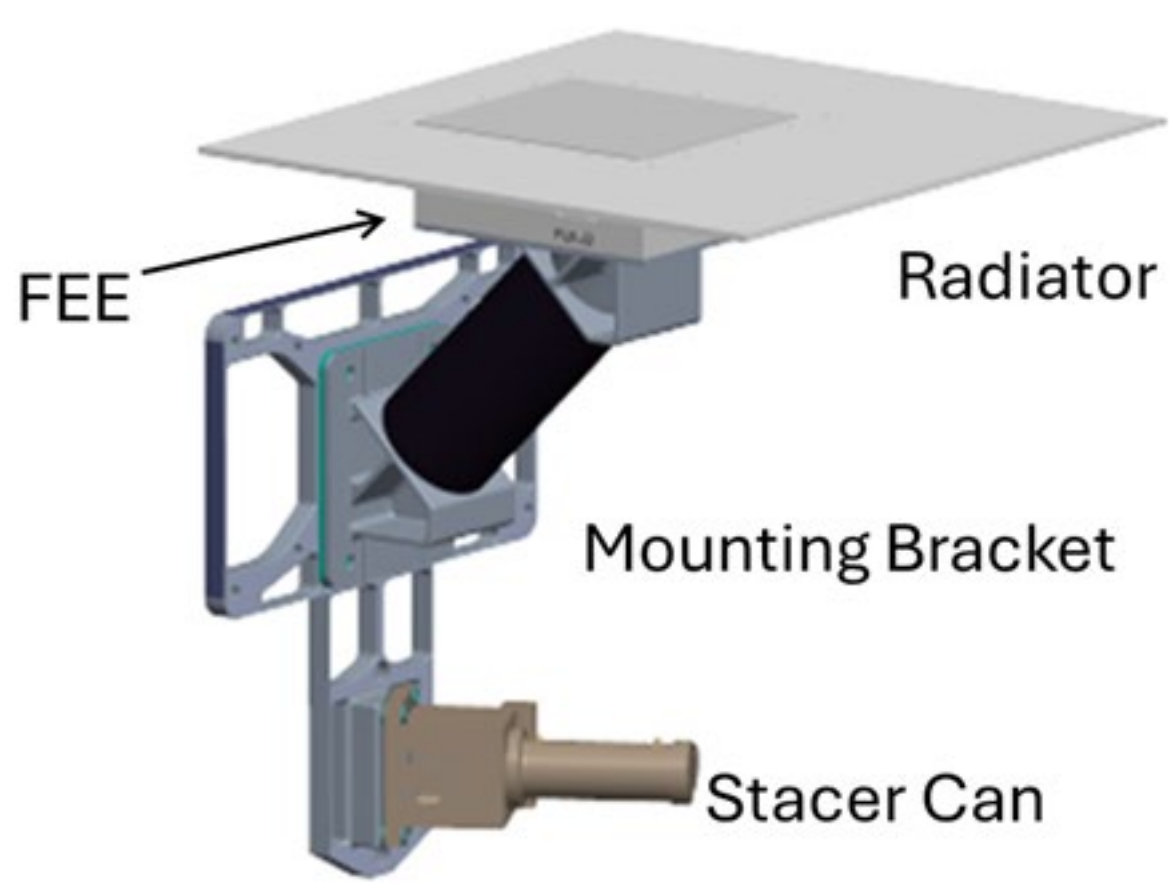

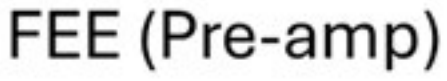

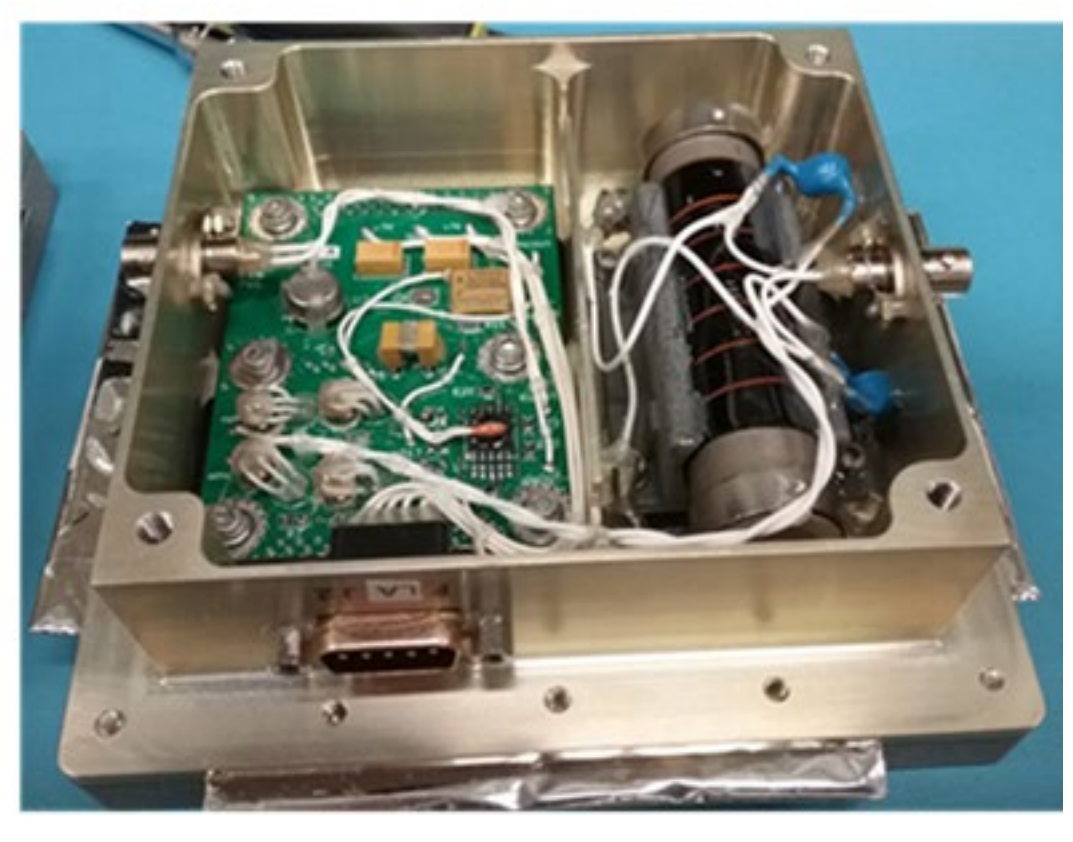

Fig. 5. (left) A solid model of the FEE assembly with radiator, mounting bracket, and the stowed STACER antenna. (right) The FEE in the open box showing the pre-amp and high voltage protection circuits. The overall size of the FEE assembly is: 40 cm (width) x 37.2 cm (height) x 45 cm (depth) in the undeployed configuration. Each pre-amp has a size of 5 cm x 5 cm x 2 cm. The total mass of all four FEEs (pre-amps) with STACER antennas, radiators and mounting brackets is 6.85 kg.

### *3.2 Front End Electronics*

The front-end electronics (FEE) assembly, designed and built at NASA/Goddard Space Flight Center, consists of a pre-amplifier, a radiator, and a mounting bracket (see Fig. 5). The FEE electronics (pre-amplifier) is located at the top of the mounting bracket, while the stowed antenna is attached to the bottom. The radiator is attached to the upper surface of the FEE box. The bracket is attached to the outer surface of the lander. Each FEE assembly is mounted on the lander such that the position of each STACER is co-linear with another one.

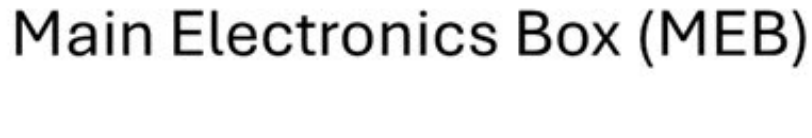

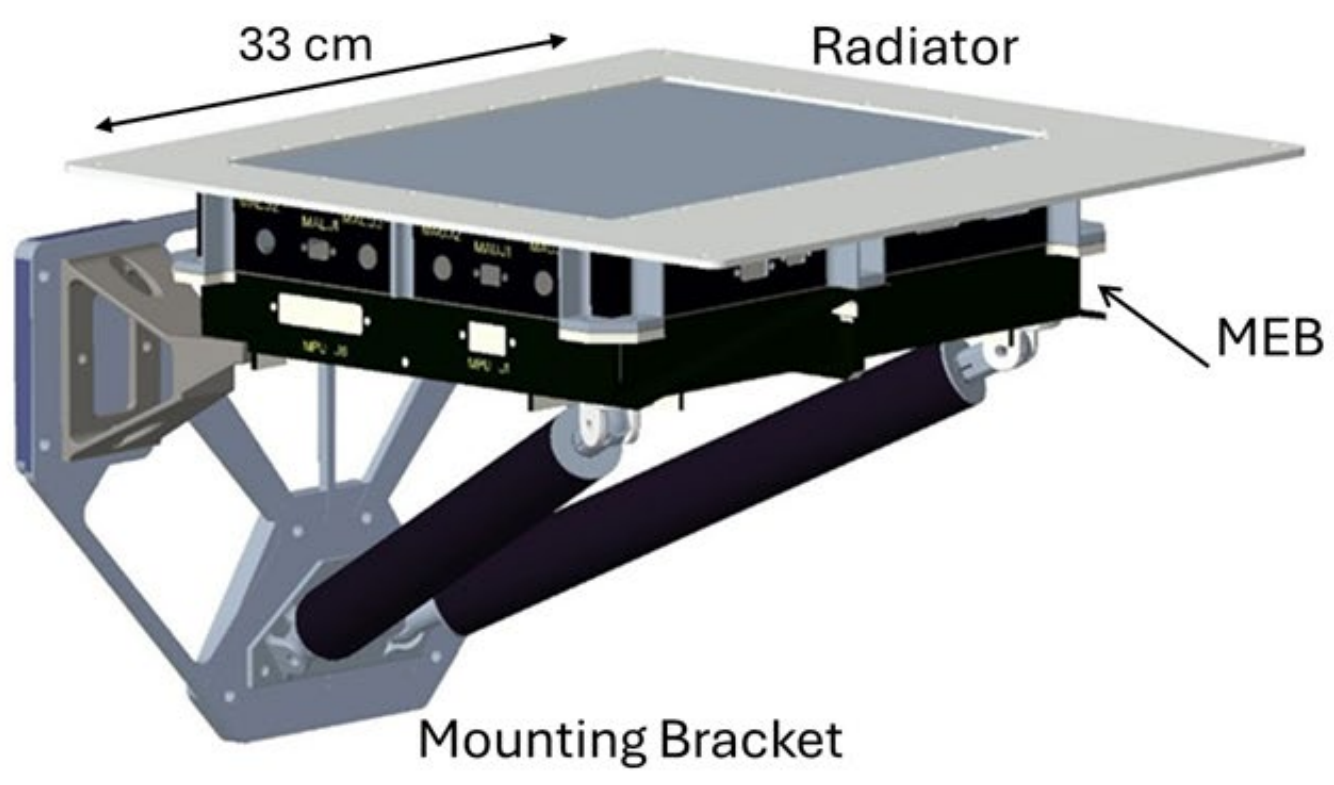

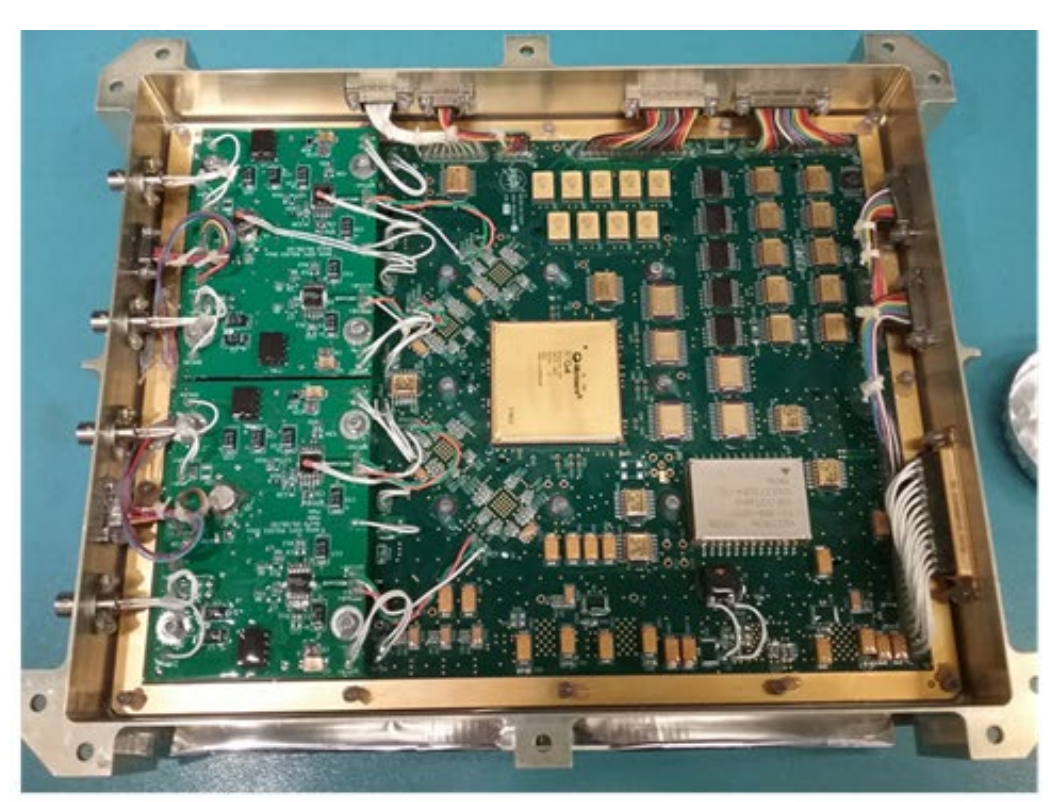

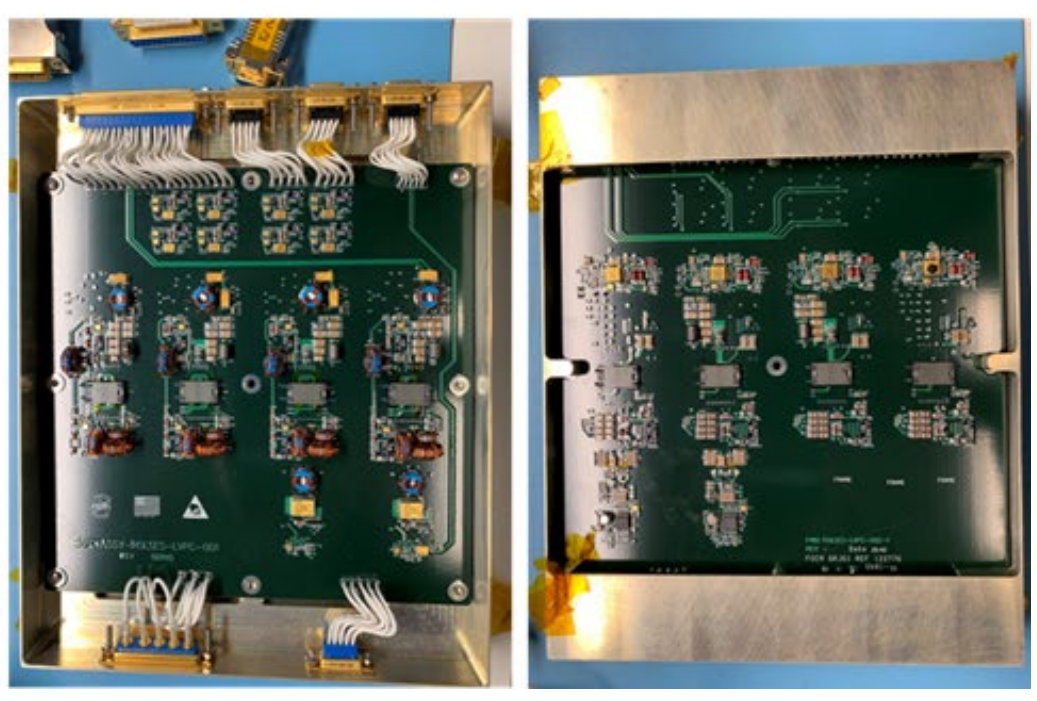

Fig. 6. (left) A solid model of the Main Electronics Box (MEB), radiator, and the mounting bracket. (top-right) One side of the Digital Unit is shown in the upper right. The digital unit consists of the main board (FPGA plus supporting electronics) and two daughter cards, which conditions the analog data for digitization. The radiator size is 33 cm x 33 cm. The height from the bottom of the bracket to the top of the MEB is 23 cm. The dimensions of the MEB itself are: 20 cm x 25 cm x 8 cm. The total mass of the MEB, radiator, and mounting bracket is 3.78 kg.

### *3.3 The Main Electronics Box*

Figure 6 shows a photograph of the MEB and the circuitry it contains. The MEB has an FPGA-driven design and layout also built at the NASA Goddard Space Flight Center. The MEB has the following subsystems: (1) two analog units (AU daughter cards), (2) the digital unit (DU) proper, and (3) the low voltage power converter (LVPC). The ROLSES team designed and manufactured all harnesses and connectors between MEB and the FEEs/antennas.

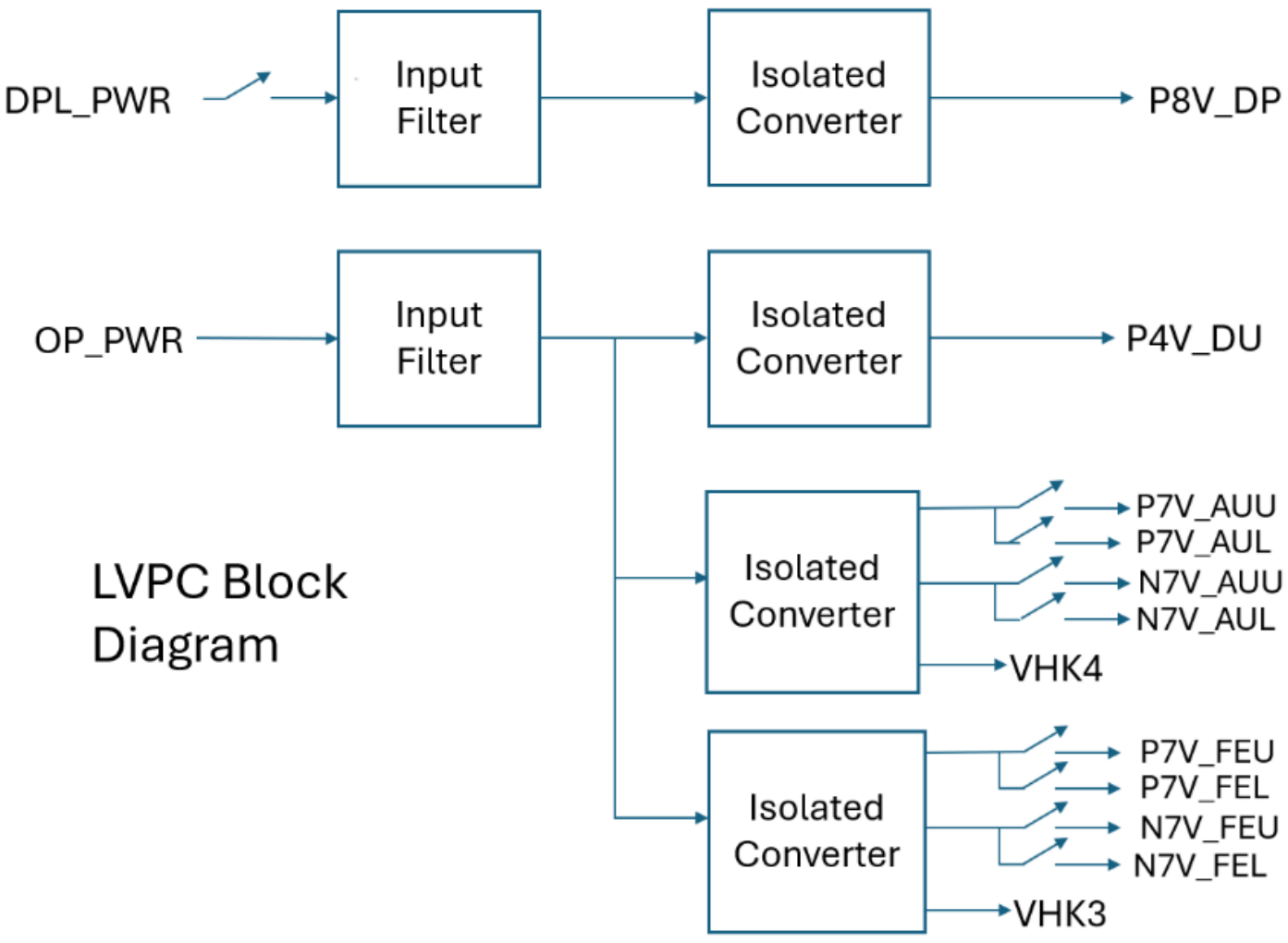

Fig. 7. The LVPC block diagram showing the two power channels and a set of isolated converters. Isolated outputs are +4 VDC for the digital units (DU), +8 VDC for deployment DPL (one time), +8 VDC for calibration CAL (all the time), +7 VDC and -7 VDC for analog units AU, and +7 VDC and -7 VDC for FEEs. P and N stand for positive and negative voltages in the diagram. VHK3 and VHK4 are housekeeping supplies for internal use of the LVPC. The average power for ROLSES is 22 W.

### *3.3.1 The Low Voltage Power Converter*

The LVPC accepts input power from the lander and distributes conditioned power according to ROLSES requirements. Fig. 7 is a block diagram of the LVPC. The lander provides two independent (+28 VDC) channels, one for antenna deployments (DPL_PWR) and the other for ROLSES operations (OP_PWR). For example, after passing through the input filter, an isolated converter produces an 8 V output for deployment. Having a second +28 VDC power feed, which is normally off, prevents any possible inadvertent antenna deployments.

### *3.3.2 The Digital Unit*

The signal flow from the antenna to the lander and the DSP algorithm as a processing chain are shown in Fig. 8. The signal from the antenna gets amplified in the FEE consisting of a JFET and a second-stage amplifier. The signal from the FEE is digitized and fed to the FFT to generate the radio spectra, which were sent to the lander. Detailed functioning of the DSP unit is shown in Fig. 8b, the digital signal from each antenna was down-sampled to high band (29 kHz to 30 MHz) and low band (1.8 kHz to 1.875 MHz). After passing through anti-aliasing filters, and polyphase filter bank, FFT was performed to generate 512 bins of spectral data (the bin size is 58.6 kHz and 3.7 kHz for the high and low bands, respectively). The accumulator block creates a rolling sum that iterates 58592 times for high band and 3662 times for the low band. Within each rolling sum, the result from the previous iteration is padded with 16 '0' bits. New 16-bit data is added to the zeros, effectively concatenating the data. After each iteration of the rolling sum, the result's 16 most significant bits are discarded, resulting in the loss of prior data. The net result of this operation is that only the last two FFTs contribute to the accumulator's output. Finally, a CORDIC algorithm (Volder, 1959) was implemented to determine the magnitude of the complex data.

The spectra in the upper and lower bands are produced separately. For the upper band, we have 512 channels in the range 0-30 MHz with a bin of ~58.6 kHz. For the lower band (0-1.875 MHz), the 512 channels have a bin size of 3.7 kHz. A single FFT in the high band (0-30 MHz) needs 1024 samples of ADC data sampled at 120 MHz, or 8.3 microseconds in the time domain. The total number of FFTs that can be performed in 1 s is 58592 in the high band. In the lower band when sampled at 7.5 MHz, we also need 8.3 microseconds for 1 FFT. Thus in the lower band (1-1.875 MHz) one can perform 3662 FFTs. In other words, the time taken to complete all the FFTs is 500 microseconds in each band. The difference in sampling rate for each band is driven by the need for higher frequency resolution in the low band. With the same number of samples per run, a lower sampling rate produces a smaller frequency resolution. Simultaneous processing of both bands would require twice as many MACC (multiply-accumulate) resources in the FPGA, which we did not have. The final design used roughly 81% of the MACC blocks in the FPGA.

The processing chain in Fig. 8b applies to all four antennas and their DSP units. The antenna labels and mapping to their DSPs are 1) LOWER A which feeds DSP A, 2) UPPER A which feeds DSP B, 3) UPPER B which feeds DSP C, and 4) LOWER B which feeds DSP D. The lander was a hexagonal cylinder that stood four meters tall and 1.57 meters wide. The antennas were mounted on four of the six sides of the lander as follows: Lower A on northeast, Lower B on northwest, upper A on south, and upper B on north faces. For convenience we denote the antennas by the corresponding DSP labels: lower A is antenna A, upper A is antenna B, upper B is antenna C, and lower B is antenna D.

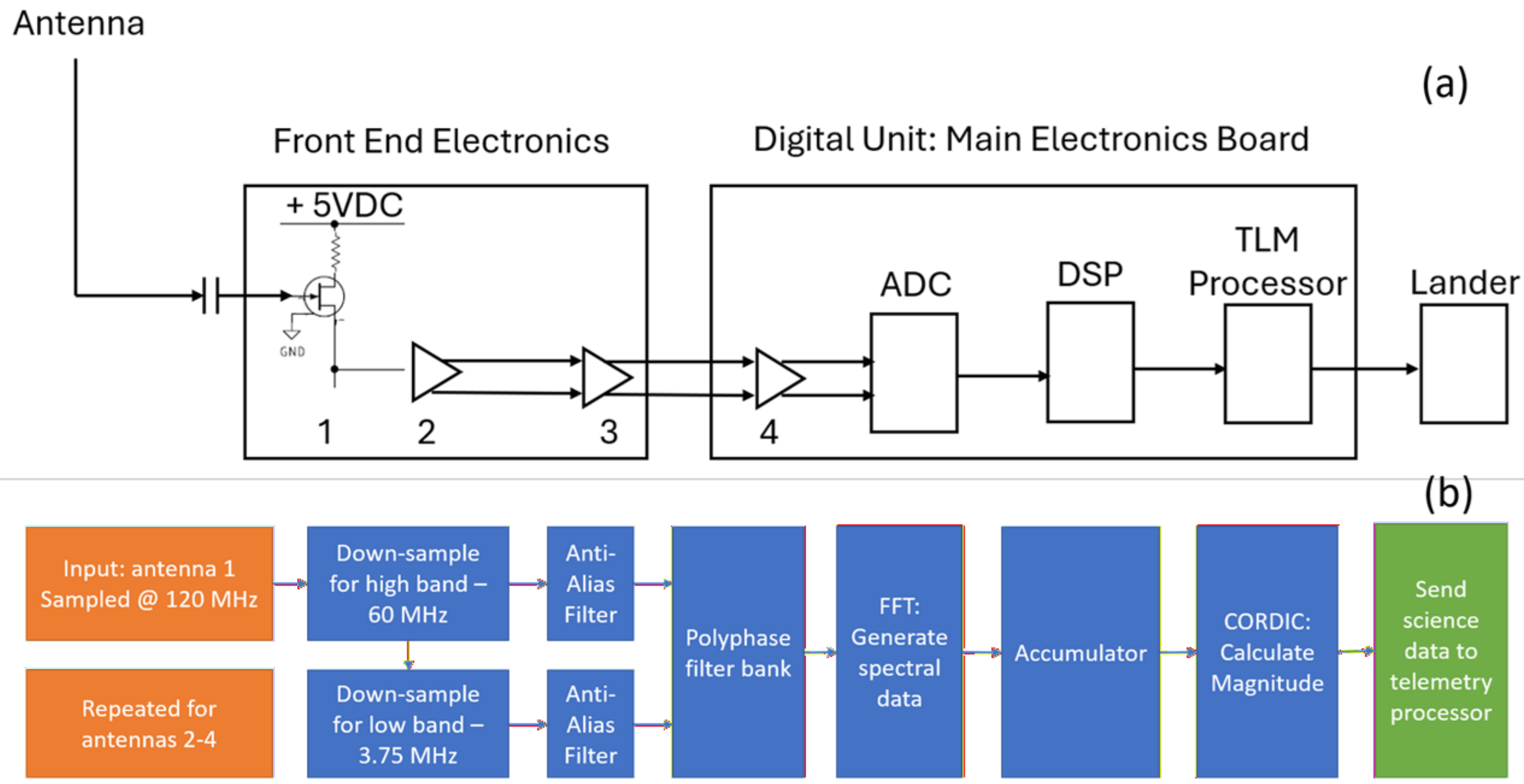

Fig. 8. (a) Signal flow for ROLSES signal processing from the antenna to the front end electronics and the Main Electronics Board and ultimately to the lander. The analog-to-digital converter (ADC) is clocked at 120 mega samples per second. The digitized signal is sent to the digital signal processing (DSP) unit whose output is sent to the telemetry (TLM) processor. ROLSES utilizes the mass storage capability of the IM-1 computer. Marked parts: 1. JFET amplifier. 2. Second-stage Amplifier. 3. Driver. 4. Receiver. (b) A block diagram of the Digital Signal Processor (DSP) function. The digital signal chain for one antenna is shown. The chains for the other antennas are identical. ROLSES has four separate and identical FFT "cores." The controls allow us to enable or shut down each core individually (e.g., for faults, save power). Each of the four streams (antenna through science packets) is treated as an independent experiment.

Post-launch review of the DSP algorithm revealed two major errors. After each iteration of the rolling sum, the result's 16 most significant bits were erroneously discarded, resulting in the loss of prior data. The net result of this operation was that only the last two FFTs contributed to the accumulator's output, and the per-second averaging of the science data was lost. Timing requirements within the FPGA were also not met, which could result in loss of functionality in certain thermal conditions.

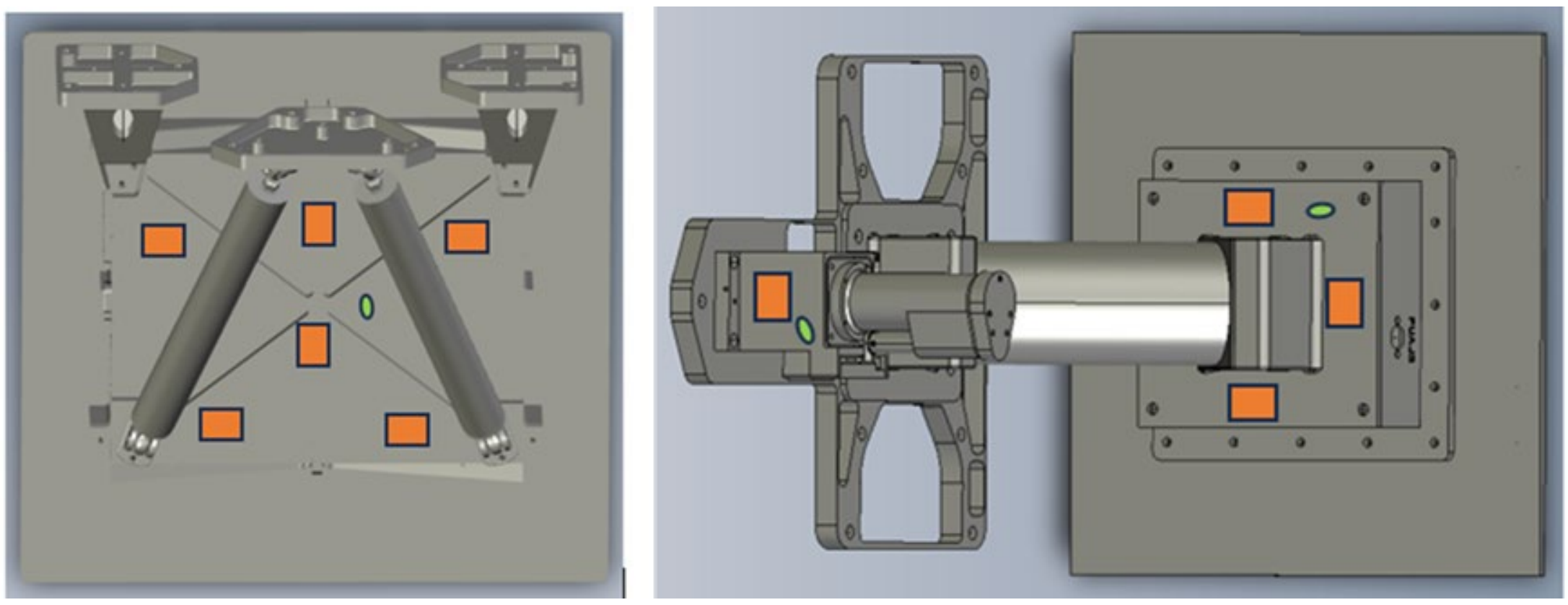

Fig. 9. Placement of survival heaters (orange rectangles) and temperature sensors (green ellipses) to the bottom of the MEB (left) and an FEE (right). In the case of an FEE, an additional heater is attached to the antenna housing along with a temperature sensor.

### *3.4 Thermal Considerations*

When ROLSES is powered off, the temperature of the MEB and FEEs need to meet Allowable Flight Temperature (AFT) limits. Therefore, survival heaters are required to meet non-operating minimum temperatures. A survival heater circuit consists of a temperature sensor and one or more Kapton heaters connected in parallel circuits. The heaters and temperature sensors (thermocouples) are attached to the bottom plates of the MEB and FEEs at selected locations (see Fig. 9). Heaters are over-taped with aluminum foil tape to spread heat out. The heaters are wired and connected to the vehicle avionics that supplied power to the heater circuits and controlled them using the flight software. The ROLSES heater resistances were sized at 27 V and a heater duty cycle of 70%.

The maximum non-op (survival) AFT limits for the MEB and FEE baseplates are 65°C and 70° C, respectively; the minimum non-op temperature is - 40°C for both. The technical specifications from the frangibolt manufacturer show an operating temperature in the range -50 °C to 70° C. At a pre-actuation temperature of 70° C or warmer, the frangibolt may inadvertently actuate. Unexpected exposure of one of the antennas to solar flux in the cruise phase did lead to an inadvertent deployment (details later). There is a trade-off: the lower the temperature of the STACER, the more energy is needed to deploy. Based on manufacturer's sample data, there was confidence that deployment over the entire temperature range would proceed without issue. In the testing phase, we were able to deploy an antenna in a Thermal Vacuum (TVAC) at – 50º C in a reasonable period of time (92 s). The spec was – 45 ºC, so the deployment was not an issue. Room temperature deployments took 30-46 s. All tests were performed with 7 VDC and 1.1-1.2 A current. We also tested another antenna with similar spec but using a lower voltage: 5.2 VDC. In this case, the deployment took longer (160 s) starting from – 50º C. Thus, the temperature range -45 ºC to +60 ºC works for cruise and deployment stages of the STACERs with plenty of margin. Once deployed, there were no special concerns for the STACERs' temperatures.

For both MEB and FEEs, survival and operational setpoints are -40 ºC (turn on) and -35 ºC (turn off) and -40 ºC (turn on) and -35 ºC (turn off), respectively (assuming duty cycle is ≤70%). The turn on and turn off refer to the heater. When the temperature falls to − 40º C (the minimum temperature in Table 3), the heater is turned on; once the temperature reaches −35º C, the heater is turned off. This action ensures that the MEB and FEEs are above the lower limits of their survival temperature and the power is not wasted. The duty cycle refers to the percentage of time when the heaters are on. The 70% duty cycle gives margin for the system: if things are colder than anticipated, there's another 30% of the time when the heaters can be turned on, up to 100% of the time if needed. The lander provides temperature monitoring, control, and heater power to maintain the ROLSES payload above the lower temperature limits in Table 3.

Table 3. ROLSES Temperature Limits

| ROLSES Element | Heater Controlled? | Non-Operational Temperature | | Operational Temperature | |
|---|---|---|---|---|---|
| | | Lower Limit (ºC) | Upper Limit (ºC) | Lower Limit (ºC) | Upper Limit (ºC) |
| MEB | Yes | – 40 | 65 | – 30 | 55 |
| FEEs | Yes | – 40 | 70 | – 30 | 60 |
| STACER Antennas | No | – 50 | 70 | – 45 | 60 |

### *3.5 Mechanical Considerations*

Mechanical work involves building enclosures, support assemblies, and radiators for the FEEs and MEB and attaching them to the lander along with the stowed STACER antennas. The radiators were made of 6061 aluminum, which is a precipitation-hardened alloy primarily composed of aluminum, magnesium, and silicon. The same material was used for FEE and MEB boxes, which also function as a radiator surface. Optical Solar Reflector (OSR) tiles for solar input mitigation were installed on MEB and FEE radiators, followed by the installation of solar reflective tapes. OSRs were made of vapor-deposited aluminum to reflect solar radiation and dissipate internal heat into deep space. A thermal blanket was fabricated to cover the ROLSES subsystems. The total mass of the ROLSES system is 13.24 kg (four FEEs: 6.85 kg; MEB: 3.78 kg, insulation: 0.50 kg, harnesses: 2.11 kg).

### *3.6 Concept of Operations*

There are two bi-directional communication buses between ROLSES and the lander's command and data handling (C&DH) subsystem. ROLSES' two-bus system simplifies the design and eliminates contention between commanding and telemetry. Commands are sent to ROLSES on the Engineering Bus and ROLSES responds. Science and housekeeping data are sent to the lander's C&DH system, and the C&DH system responds. Each packet can be routed to the C&DH's mass store and/or the real-time downlink channel; ROLSES's real-time allocation is 1 kbps.

ROLSES has no internal storage, so science and housekeeping data are stored in the lander's C&DH. Antenna deployment is performed using real time commands, one antenna at a time. Once commissioned, the instrument operation is almost autonomous with little to no commanding. Light commanding is required when approaching and moving into lunar night to enhance the chance of observing more dust impacts. This capability was not used as a result of the lander's troubles. The time synchronization message sent by the lander to ROLSES is in a "mission elapsed time" (MET) format with the least significant bit equal to 1 second. ROLSES uses the pulse per second (PPS) discrete signal provided by the lander for time stamping data (ROLSES maintains its own MET based on its own Temperature Compensated Crystal Oscillator (TCXO); both METs are included in every packet). The PPS signal (TCXO) has the accuracy specification of 1 ppm.

#### 3.6.1. Programming ROLSES

The ROLSES architecture is powerful, flexible, and easy to "program." The machine can be programmed in a number of ways: a) what operations are performed and when; b) configuring telemetry; and c) programming the DSP machine. ROLSES' major frame lasts 8 seconds and is subdivided into eight 1-second microframes. Each microframe includes operations such as launching DSP engines for either high or low band, acquiring raw data, or collect housekeeping data (see the command matrix in Fig.10). Multiple operations can take place in a signal microframe and the fine timing is designed such that collisions will not occur during normal programming. The operators simply put check marks in the correct boxes and a single command uploads this table to ROLSES.

| MICRO FRAME NUMBER | 7 | 6 | 5 | 4 | 3 | 2 | 1 | 0 | |
|---|---|---|---|---|---|---|---|---|---|
| HK ADC | X | X | X | X | X | X | X | X | F C |
| GO LOW DSP | | X | | | | X | | | F C |
| GO HIGH DSP | | | | X | | | | X | F C |
| DUMP LOW DSP | | X | | | | X | | | F C |
| DUMP HIGH DSP | | | | X | | | | X | F C |
| SPARE 0 | X | X | X | X | X | X | X | X | F C |
| GO TLM URGENT | | | | X | | | | X | F C |
| GO TLM HK | X | | | | | | | | F C |
| GO TLM DSP DATA | X | | X | | X | | X | | F C |
| GO CAPTURE RAW DATA | | | | | | | | X | F C |
| GO TLM RAW DATA | | | | | | | | | F C |

Fig. 10. Operator's interface for programming ROLSES – the command matrix. ROLSES works on an 8 second major frame and a 1-second microframe. There are 10 different actions that can be invoked. An activity can be scheduled to occur in a particular minor frame by placing an "X" in the programming matrix. Fine timing is controlled by ROLSES' control logic, simplifying the programming task. A single command sends the command matrix to ROLSES. The buttons F ad C at the right stand for "full" and "clear" to make configuring the matrix easier for the operator.

ROLSES' Telemetry Control offers a number of options. Each packet type can be routed to the real-time bus (1 kbps for ROLSES) and/or the lander's main store. While nominal operation rotates the telemetry through all 4 channels, Telemetry Control has the ability to dwell on a signal channel (see Fig.11). There are three types of packets sent to the lander via the science channel: housekeeping, histograms (spectra), and raw data dumps. For the housekeeping data, there are two different telemetry data points: priority and regular. During initial development, we anticipated that there might be different information from the "priority" telemetry points and regular "housekeeping" telemetry points. The "priority" package is for near-real-time operations (e.g., deploying the antennas), within about a minute. By default, the housekeeping packages are sent to the lander's mass store, which are then sent to the ground within ~ 24 hours. Although redundant, the housekeeping packages had guaranteed delivery, while the priority packages did not have guaranteed delivery. It turned out that the system works fine with both types of packets, and priority and housekeeping packages contain identical information. Each packet has two routing bits. A packet can be stored in the lander's mass store and/or can be sent to the ground via the real-time (stream). Default routing is shown in Fig.11. By deselecting both "store" and "stream" packets can be inhibited.

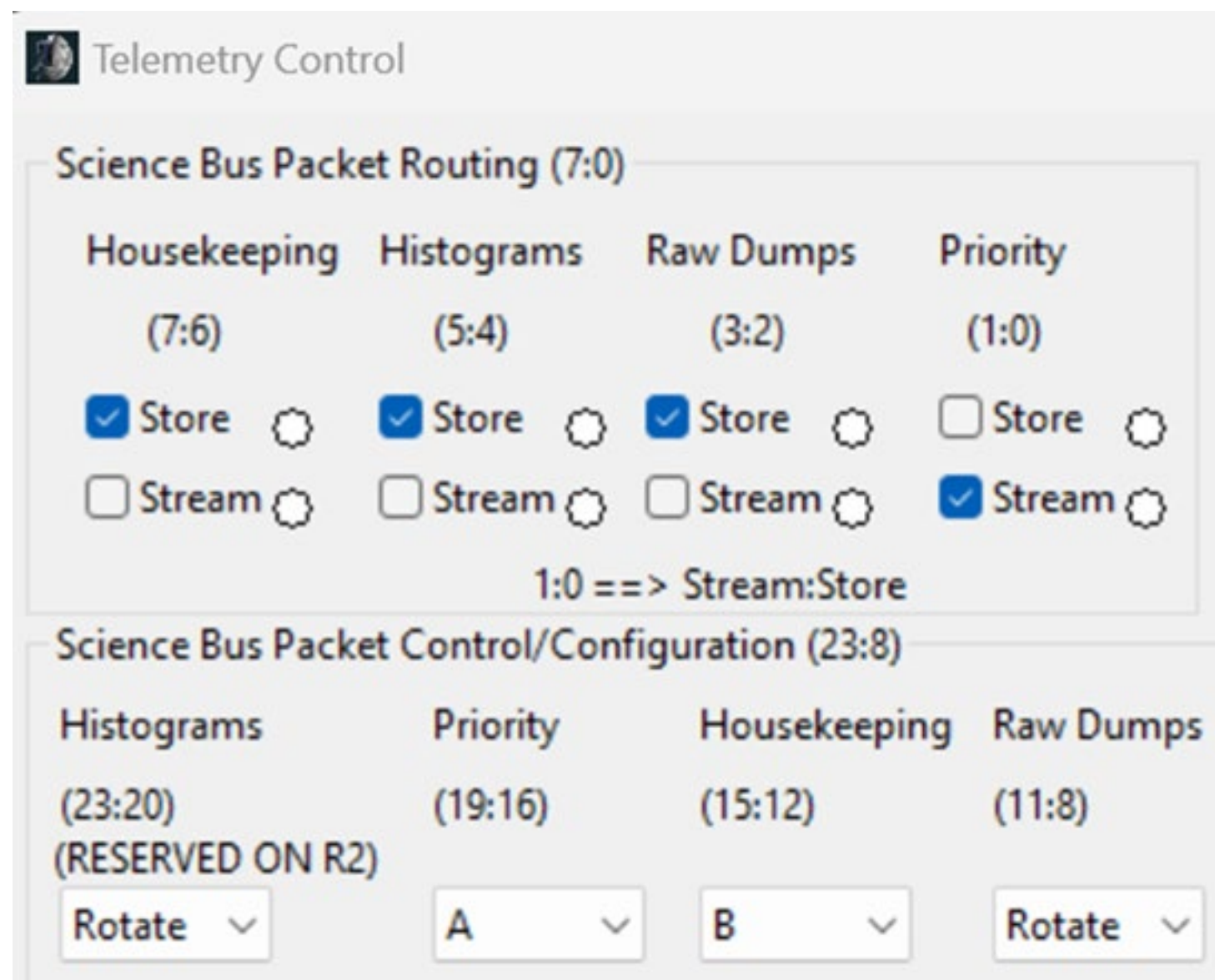

Fig. 11. Operator's interface for programming telemetry control. ROLSES' telemetry can be programmed with a simple interface. Options include routing of packets to either real-time downlink and/or mass storage. Downlink bandwidth can be controlled to either rotate through all channels or dwell on a single channel. To preserve downlink resources, certain data products can be inhibited. A single command sends the telemetry control matrix to ROLSES.

During nominal operations, packets to be transmitted are rotated. For example, since there are four antennas and processing channels, during "rotate" all channels, A, B, C, and D will be sent to the lander in turn. Optionally, a "dwell" function may be enabled. A dwell can be used to focus on one packet type. For example, during antenna deployment operations, only one "flavor" of housekeeping data is transmitted, giving higher time resolution for these critical operations.

The DSP engines are also configurable. Each channel can be programmed for the number of FFT blocks, for example. Other options are to be "off" (power commands can hold this DSP logic in reset), "run," or exercise a test mode using a numerically controlled oscillator. For this ROLSES mission, the number of FFT blocks remained fixed (58,592[a]) and the other options were not used (see Fig. 11).

[a] A logic error resulted in the DSP processing one FFT for each histogram.

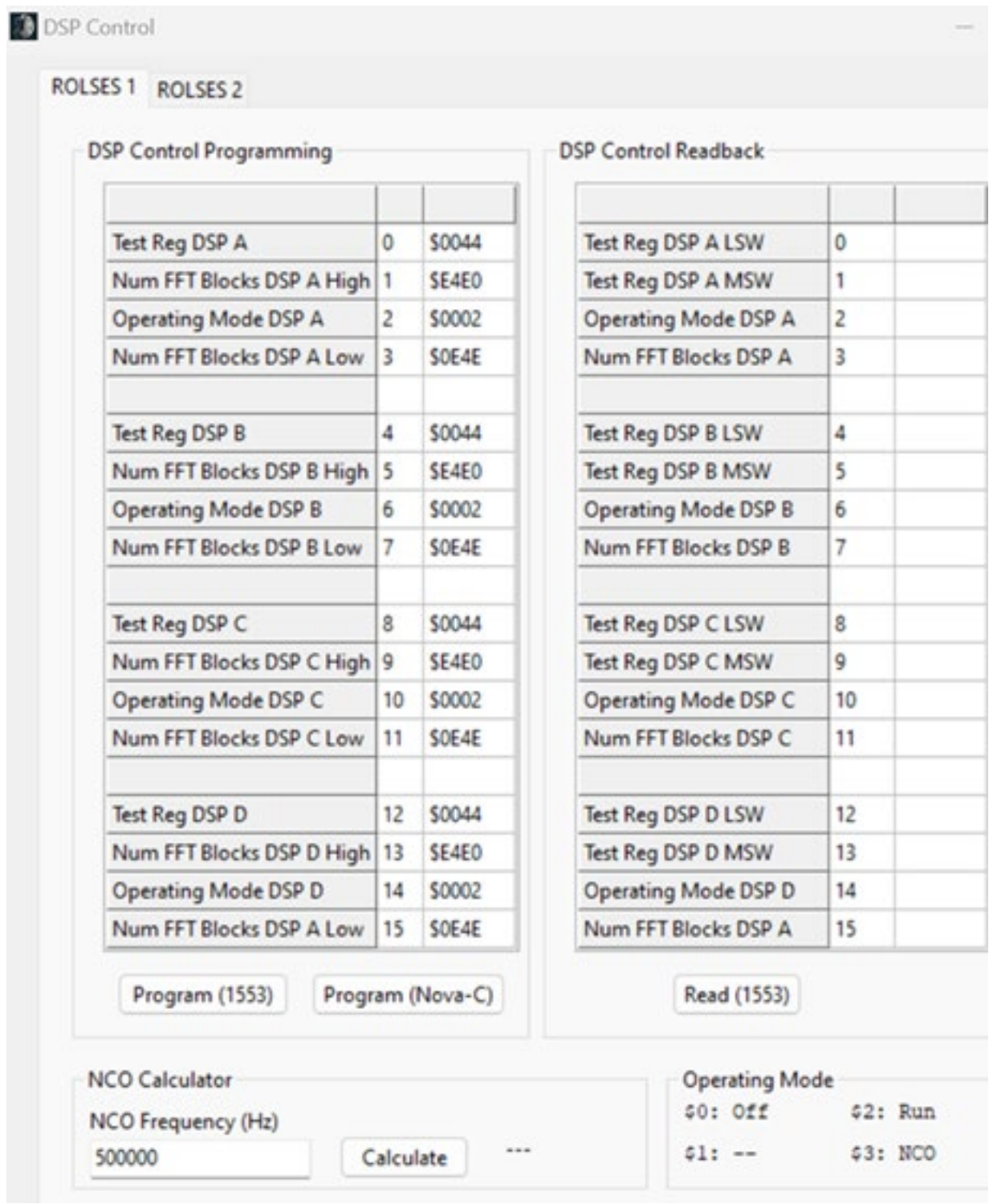

Fig. 12: Operator's interface for programming DSP control. By entry into the table shown above, the number of FFTs for a calculation can be programmed for both high and low processing bands, trading off power for quality of results. The DSP operating modes are also controlled by this table. Modes are RUN, OFF, and TEST. A single command sends this command matrix to ROLSES.

### 3.6.2 Signal Statistics

ROLSES data is sent to the lander either in processed (passed through DSP) or unprocessed (raw). ROLSES also contains an on-board signal statistics module. The module listens in to the output of the ADCs, which is sent to one of the raw memories and/or one of the DSP channels. Statistics include min/max values for each 1 second period, a running count of min/max values, an average, and an ADC out-of-range counter. Along with the health of the hardware, these values are useful for setting gains for each channel. As a result of the truncated mission, no gain changes were made. The low gain mode was functional, and no out-of-range signals were detected (see Fig. 13).

Signal Statistics

| | Min (s) | Max (s) | Max-Min (s) | Min | Max | Total (8s) | Average | OOR Total (8s) | --- |
|---|---|---|---|---|---|---|---|---|---|
| ADC_A | | | | | | | | | --- |
| ADC_B | | | | | | | | | --- |
| ADC_C | | | | | | | | | --- |
| ADC_D | | | | | | | | | --- |

Reset Stats | Read | Monitor | Dump to Logfile | Comment ---

For Total and Out of Range Total, 1,024 samples are collected every 1 s and accummulated over 8 s.
Average is updated every 8 s.

Fig. 13. Real-time Signal Statistics. The ROLSES hardware samples the incoming raw data stream, processes the data, and produces a variety of statistics including, minimum, maximum, total, and ADC OUT-OF-RANGE. These statistics are both stored in the lander and downlinked on the real-time channel, giving engineers an insight into the health of the hardware and the signal environment.

### *3.7 Communication Architecture*

ROLSES operations are performed via an internet protocol security (IP sec) tunnel connecting ROLSES web server to the lander control system as shown in Fig. 14. The ROLSES web server provides data to the users. ROLSES web server helps make three types of communication with the lander control system: 1. Real time command upload to ROLSES, 2. Download streaming data, and 3. Download moderate size data files stored in lander C&DH/IT system.

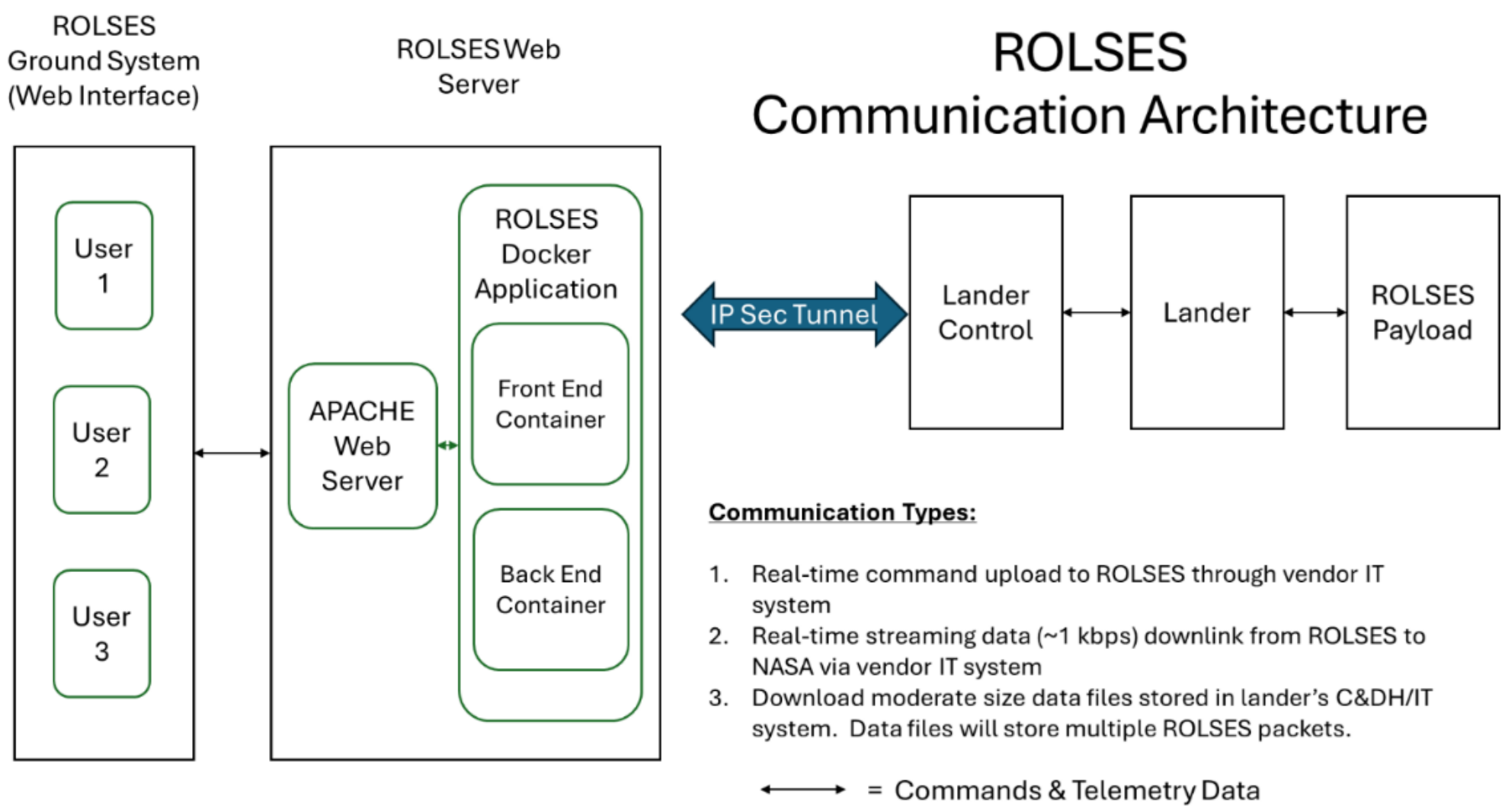

Fig. 14. A schematic of the ROLSES communication architecture showing the interface between the lander control and the ROLSES web server via the Internet Protocol Secure Tunnel (IP Sec Tunnel)

## 4. Launch and Landing

Following launch, the IM-1 mission followed a Trans-Lunar Injection (TLI) trajectory and took slightly more than 6.5 days to land. The baseline attitude was that the bottom of the lander was pointed to the Sun ("tail-to-Sun"). This was to protect the cryogenic tanks containing liquid methane and liquid oxygen from overheating. The orientation also shielded sensitive instruments on the top deck, which pointed away from the Sun. The spacecraft was in a "tail-to-Sun" orientation until the final trajectory correction maneuver (TCM) that occurred on 2024 February 20 at 21:07 UT. After this, there was no shadowing in the changed Sun-spacecraft angle. The housing of antenna D was facing the Sun and received excess solar flux for a few hours resulting in the deployment and damage of the antenna at 01:12 UT on 2024 February 21 when the lander was ~9200 km from the Moon. ROLSES was powered on to make use of the deployed antenna and observations were made during 03:17 – 04:37 UT on 2024 February 21. After landing at 23:23 UT on 2024 February 22, it took a few days to power on ROLSES using low gain antennas because the high gain antenna could not be used due to the off-nominal resting position of the lander. When ROLSES was powered on, at 16:49 UT on 2024 February 26, telemetry indicated that antenna B had deployed sometime between landing and turning on ROLSES. Since the south – southwest panel of the lander was facing the lunar surface, antenna D mounted to the panel to the left and the antenna B mounted to the panel on the right were lying on or were directed into the lunar surface. Cameras on board the lander captured images of the deployed antenna D in the cruise phase. Figure 16 shows one of these pictures in which the deployed antenna D can be seen. In other pictures, the tip of the antenna D was pointing in different directions at different times indicating some damage to the antenna. However, the antenna was electrically intact and obtained data for ~83 min in the cruise phase.

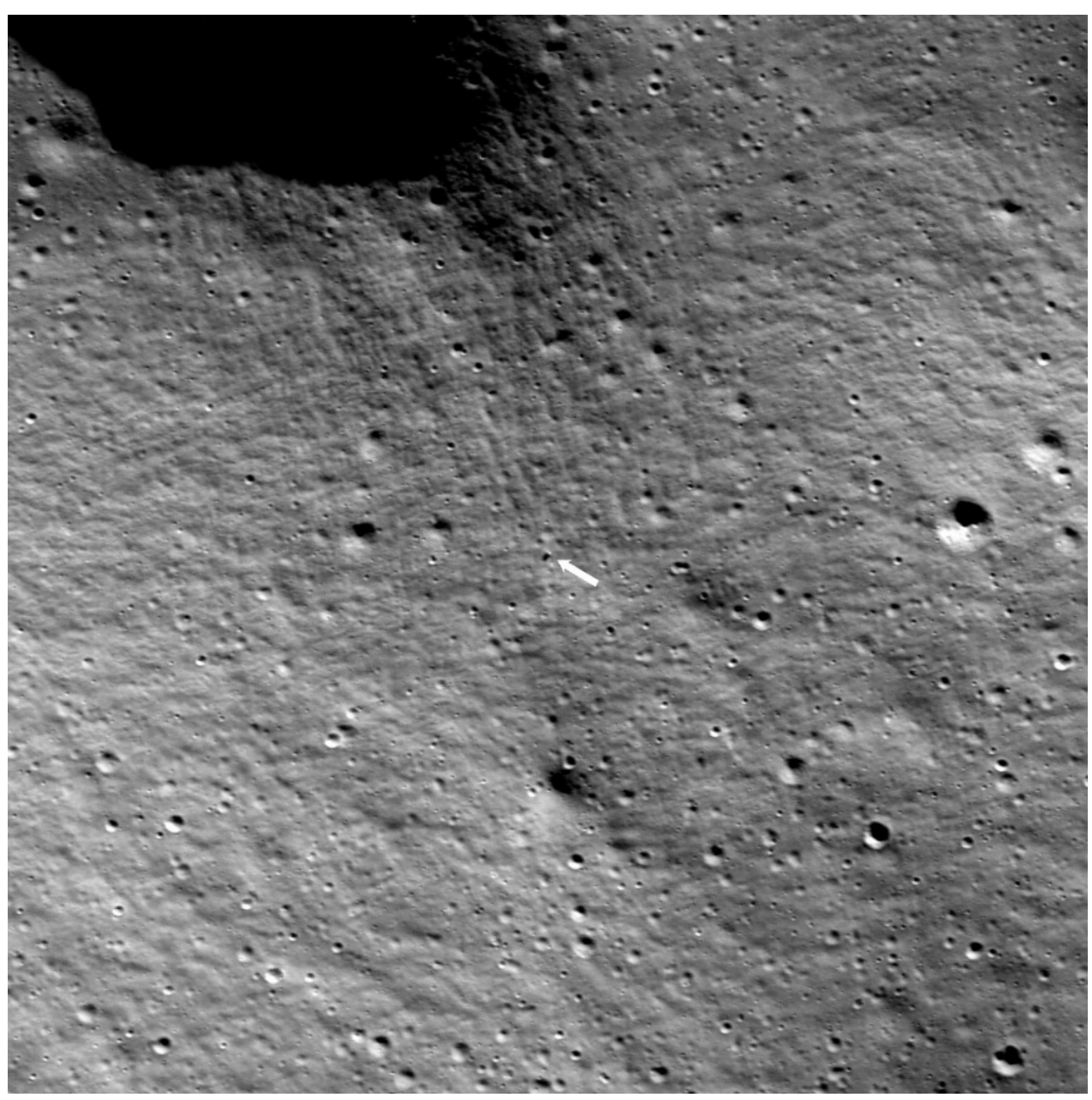

Fig. 15. Landing site of IM-1 mission (pointed by white arrow) revealed by NASA's Lunar Reconnaissance Orbiter (LRO) image taken at 06:57 UT on February 24, 2024 (-80.13, 1.44) by the LRO Camera: Narrow Angle Camera, NAC). degrees east longitude, at an elevation of 8,461 feet (2,579 meters). The image is 0.973 km wide, and lunar north is to the top. The landing location was ~1.5 km from the target location Malapert A (-80.20594, 0.93919). (LROC NAC frame M1463440322L, credit: NASA/Goddard/Arizona State University.

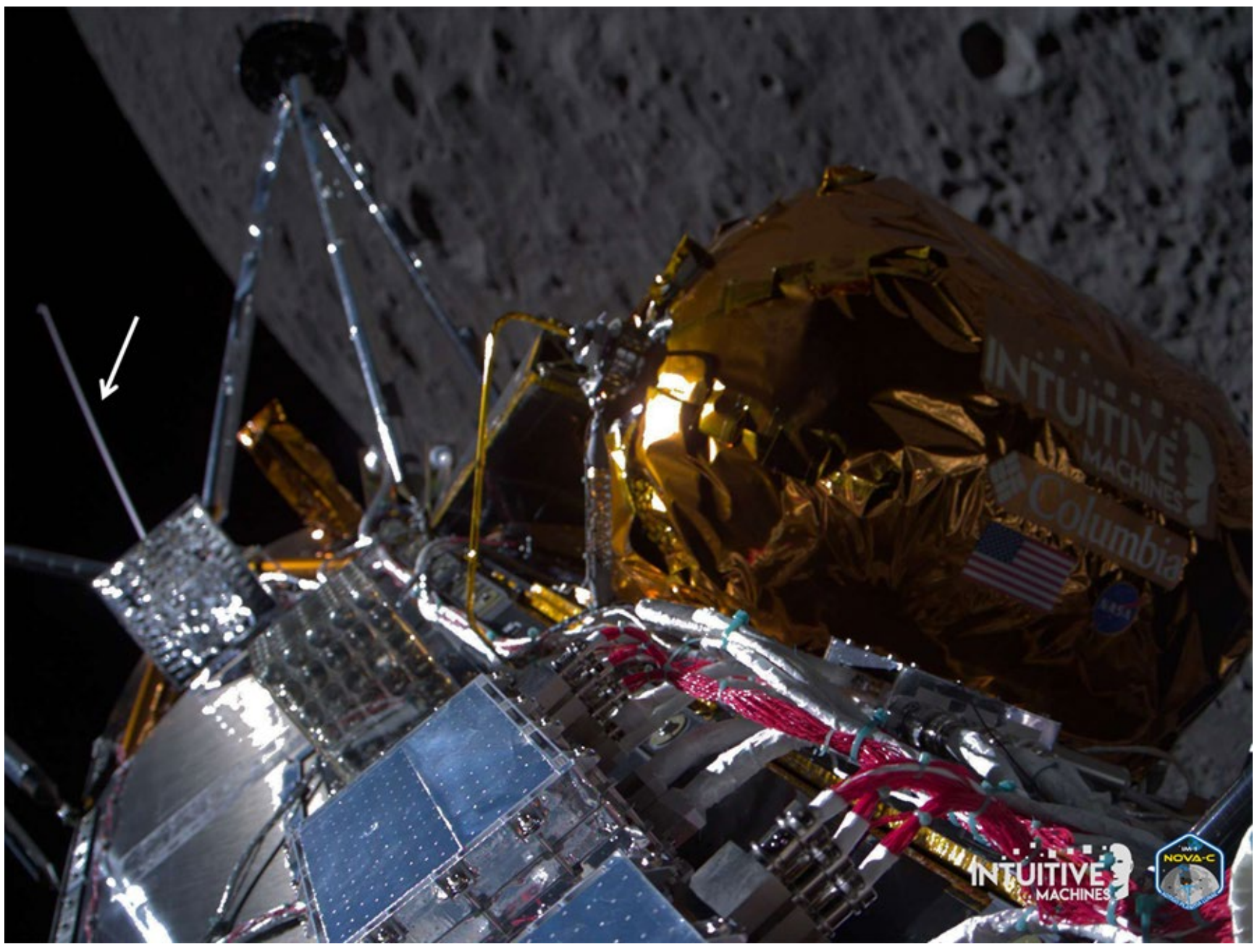

Fig. 16 Antenna D that inadvertently deployed in transit (pointed by arrow). In other pictures show that the antenna was pointed at different directions at different times indicating it was damaged but electrically connected so it was possible to make observations.

The remaining two antennas (A and C) were deployed by command. ROLSES was instrumented to track STACER temperatures and the times from "firing" the antenna until the antenna was deployed. To deploy an antenna, the 8 VDC power supply was turned on and then send a signal to an FET. With the power supply on and the FET turned on, there would be a current draw of about 1.2 amps. Figure 17 shows the current and temperature profiles of the two antennas (A and C) deployed. As noted before, the deployment time depended on the pre-actuation temperature of the antennas. Antenna A was at 48 °C when the deployment started, so the time taken to complete deployment was ~20 seconds (mission elapsed time, MET 693 s to 713 s). On the other hand, Antenna C was at a much lower temperature (-5 °C) in the pre-actuation phase, so the deployment time was longer (48 seconds, MET 1073 s to 1121 s). After deployment, it was estimated that antenna C pointed at ~45º from zenith and antenna A pointed near zenith Due to the tipping over of the lander, commanding of the ROLSES instrumentation was limited and there was no opportunity to command ROLSES' signal chain to high gain (voltage gain 30 dB). Therefore, the gain setting was at its lowest (voltage gain 10 dB) for all four antennas for all data intervals, and the performance of the instrument could not be optimized to extract remote signals. While on the lunar surface, only 36 minutes of spectral data was obtained compared to the projected 170 to 340 hours of data. In this paper, we present results of analyzing this 83 min of data, while more detailed analysis of the 36 minutes of surface data has been reported elsewhere (Hibbard et al. 2026).

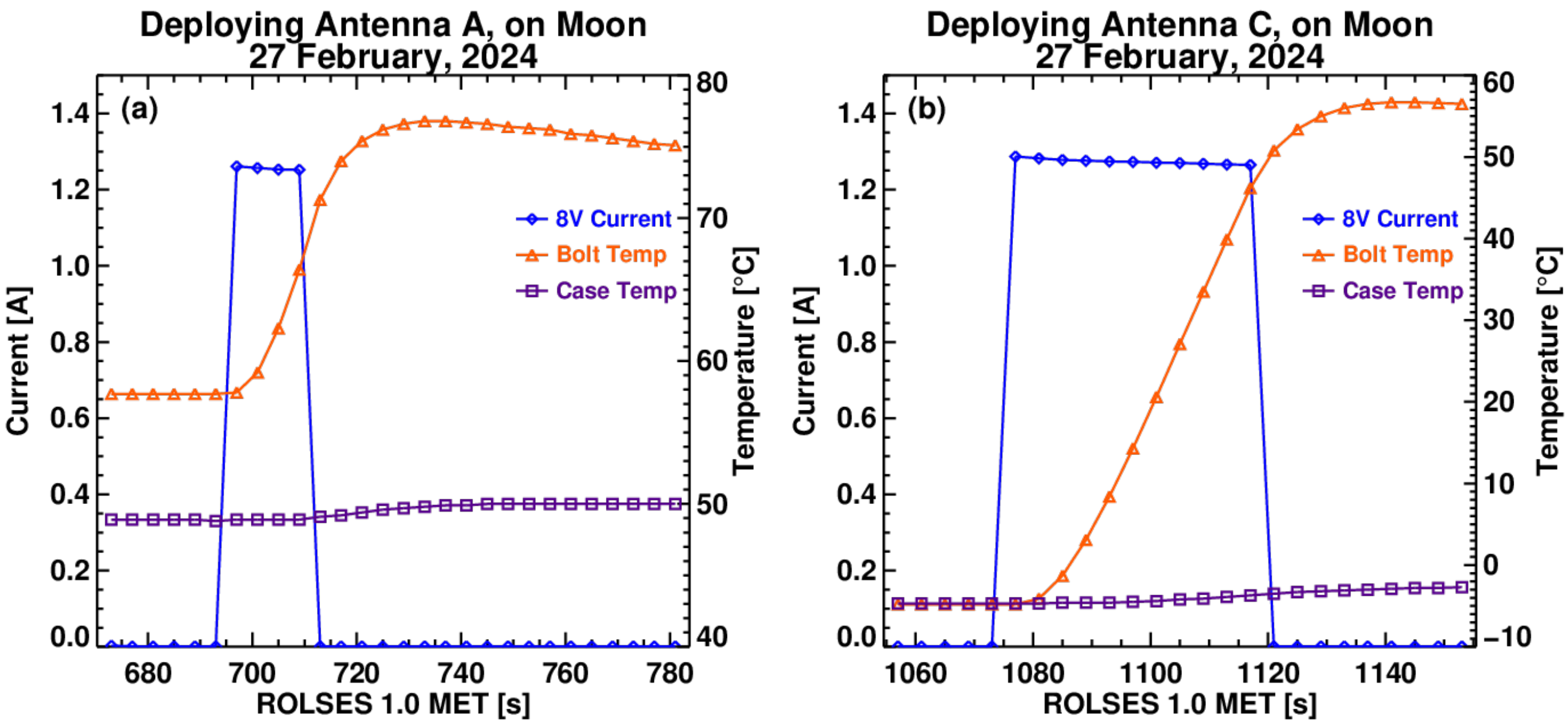

Fig. 17. Time profiles of the deployment current (left Y-axis) and temperature (right Y-axis) in the frangibolt of the two antennas deployed by command: (a) antenna A and (b) antenna C. The temperature of the antenna housing (Case Temp) is shown for reference. The time is the mission elapsed time (MET) in seconds. The data were obtained from telemetry and indicate that the deployments were nominal. Each antenna has a "beginning of deployment switch." It's a single bit for each antenna in the housekeeping packet so we can see if it is deployed or not.

## 5. Data Processing

ROLSES data processing is described in the flow chart (Fig. 18). ROLSES telemetry packets are stored on-line on a password protected Intuitive Machine's account. This data is composed of housekeeping data, histogram (spectra science) data, and raw (waveform science) data packets. The data are manually transferred to a NASA laptop where the science packets are extracted using a code derived from the data packet descriptions given in the ROLSES Command and Telemetry Dictionary. The science data are converted into Level 1 (L1, non-calibrated) science data, and Level 2 (L2, calibrated) science data in common data format (CDF). Note that no preflight calibration files were made of the raw waveform data, so that data is not calibrated. The science L1 and L2 data were then copied to a password protected NASA Box account where it can be retrieved by the ROLSES Science and Engineering Teams and by the Planetary Data System (PDS). The PDS has made the data public (https://pds-ppi.igpp.ucla.edu/bundle/urn:nasa:pds:clps_to2_im_rolses#:~:text=CLPS%20TO2%2DIM%20ROLSES%20Bundle,New%20Release, Gopalswamy and Boardsen 2025). The datasets are described in the following paragraph.

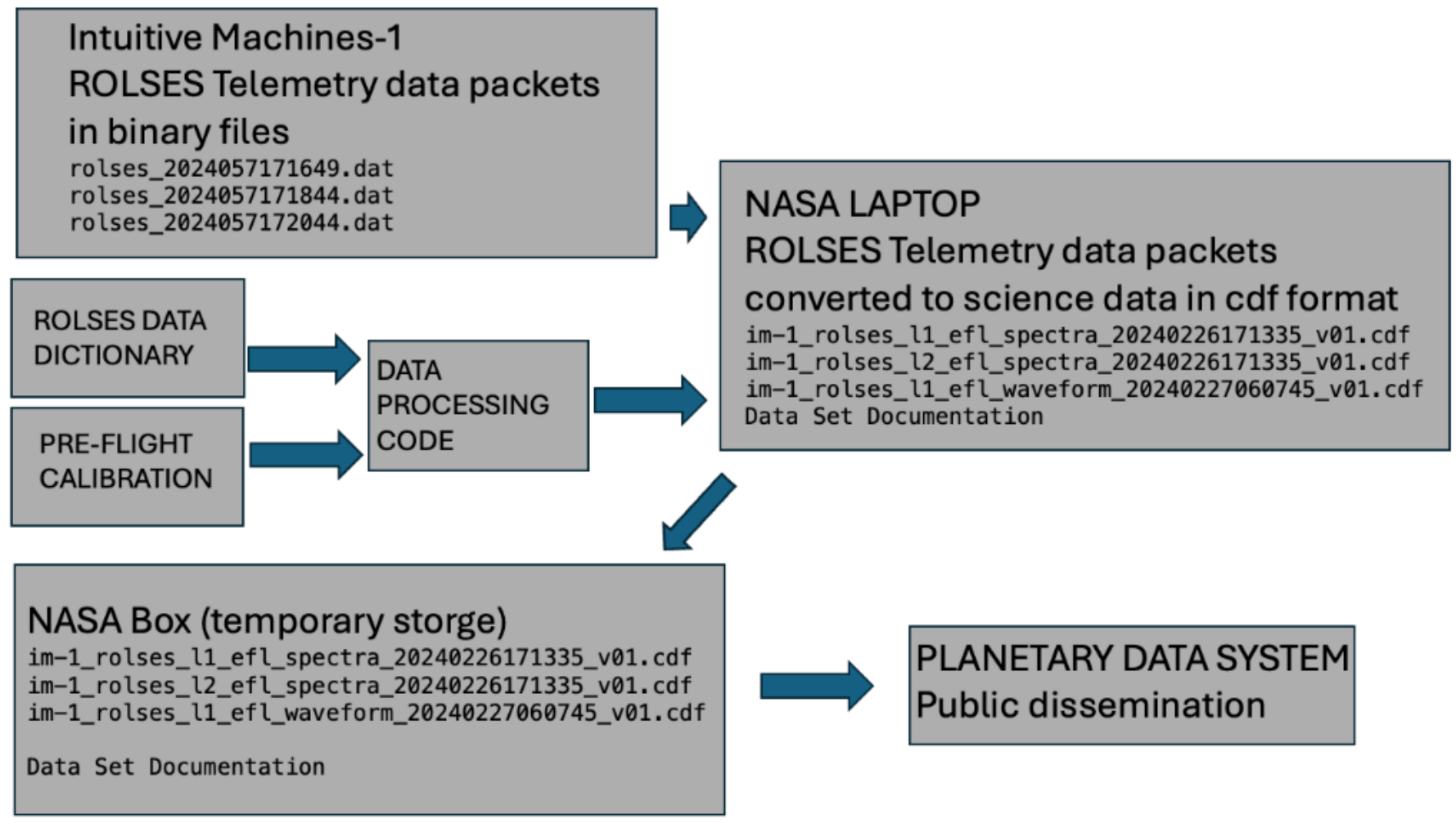

Fig. 18. Data processing flow diagram.

The digitized signals, sampled at a rate of 120 mega samples per second, are fed into Digital Signal Processors (DSP) for spectral analysis. Each antenna has its own dedicated DSP where the signals are processed into spectrum for both low (~0.003 to ~1.9 MHz) and high frequency bands (~0.06 to ~30 MHz) of 512 frequency channels each. The cadence of both bands from all 4 antennas (8 spectra) is 8 seconds, with the low band and high band staggered by 4 seconds. No cross-correlation was made between DSP's outputs so only the absolute values of the fast Fourier transform (FFT) amplitudes are returned. Besides spectra (often referred to as histogram data), the raw digitized data may be returned (raw waveform). ROLSES' operations (low band, high band, and raw; dwelling on a channel or rotating; or disabling channels) can be configured by simple table uploads or commands as described in Figs. 10-11.

The dataset variable names contain letters like 'DH' where 'D' refers to DSP D and 'H' for high frequency band, 'AL' where 'A' refers to DSP A and 'L' for low frequency band, etc. The uncalibrated level 1 (L1) data consists of uncalibrated FFT absolute amplitude from the DSPs. The calibrated L2 data consists of calibrated FFT absolute amplitude from the DSPs in volts, along with power spectral densities where 2.5/4 m is used as the effective antenna length. Note for ease of intercomparison the power spectral densities (PSDs) are computed regardless of whether the antenna is deployed. The ground test data is included because the voltage levels are almost identical to those observed during cruise and on the lunar surface. The raw waveform voltage is uncalibrated, and it is dominated by strong power supply spikes and of instrument/spacecraft generated electromagnetic interference, which also dominated the DSP signal processing.

The archive of spectral data consists of ground integration on December 7, 2023, for 3.3 minutes, cruise phase on February 21, 2024 for 83 minutes, Lunar 1 on February 26, 2024, for 8 minutes, and Lunar 2 on February 27, 2024 for 28 minutes of data. The raw waveform data archive consists of ground integration on December 7, 2023 consisting of 35 to 36 waveforms per antenna, and Lunar 2 on February 27, 2024 consisting of 27 to 28 waveforms per antenna.

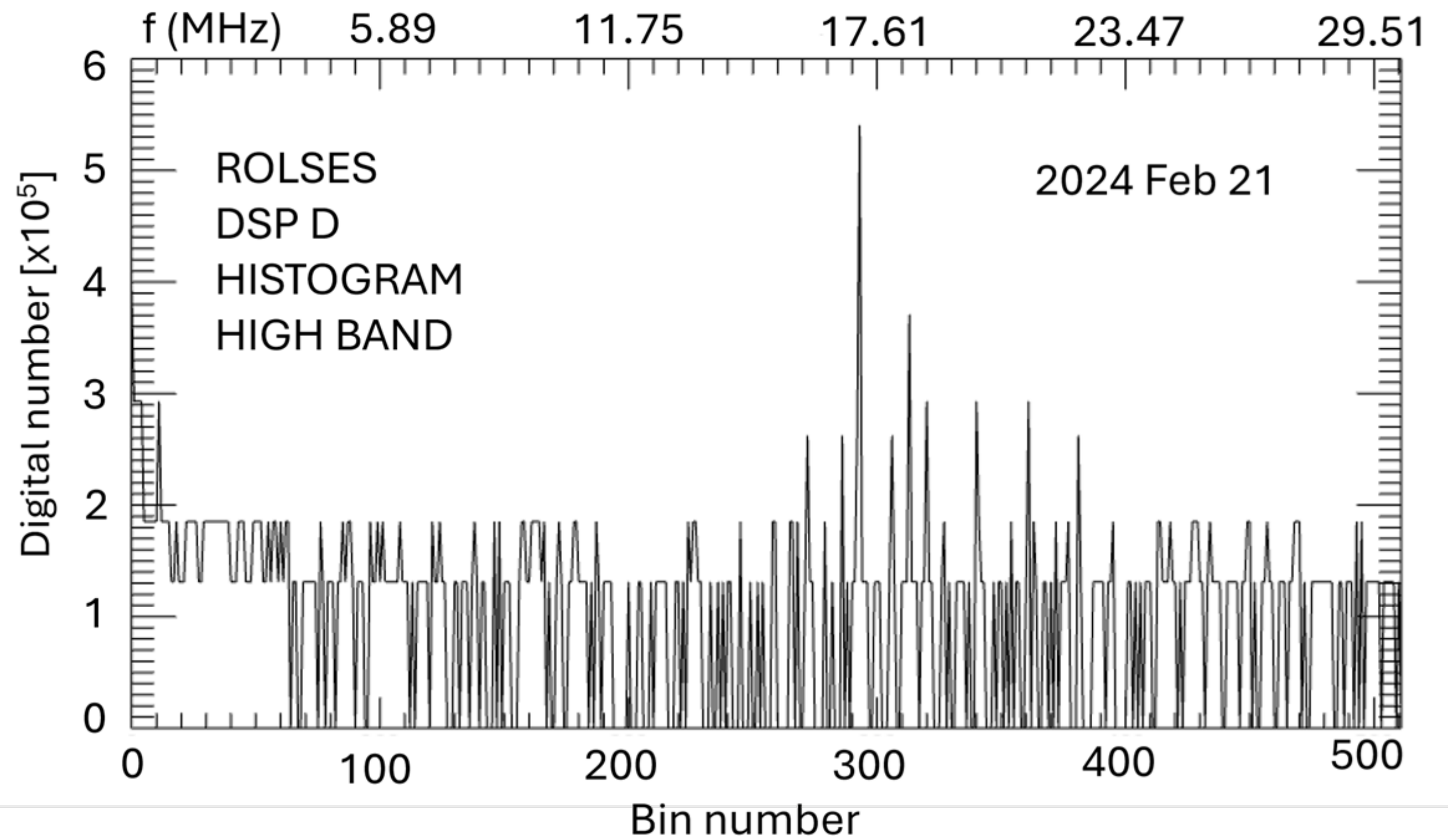

Fig. 19. A spectrum out of DSP D connected to the antenna that deployed in transit. The bin numbers on the X-axis represent the 512 channels of the spectrometer. Each bin has a size of 58.594 kHz. The channel frequencies are k*58.594 + 29.3 kHz with k = 0-511. The zeroth bin has a frequency of 29.3 kHz, while the 511th bin has a frequency of 29970.8 kHz. The Y-axis is the digital number (DN) in arbitrary units. When no significant signal is input to the DSP, the output is two constant noise levels at DN = 131,072 and DN = 185,364 known as 'shelves' and are an artifact of the DSP design implementation.

## 6. Initial Results

Figure 19 shows an uncalibrated histogram obtained on February 21, 2024 when one of the antennas deployed in transit due to excess solar heating. The DSP unit's outputs are spectra like this recorded every 4 s in each of the high and low bands. The constant noise levels at digital numbers (DN) around 131,000 and 185,000 are indicative of lack of signal. Tests and simulations of the DSP functioning show these "noise shelves" output from the DSP unit when no input signal is fed to the DSPs. In fact, the eight spectra obtained on 2023 December 7 with undeployed antennas shows similar shelves. The appearance of the noise shelves is a result of an error in the accumulator that did not allow averaging over 58,592 FFTs in each second. The average of 58,592 FFTs would have reduced the channel-to-channel deviations from the quantized shelf levels. Instead, only the last FFT in each second was processed. There are nine bins in the range 270 to 380 in Fig. 19 that have values well above these noise shelves and hence are real signals. The frequencies corresponding to the significant bins are in the range 16 to 22 MHz. The most intense bin (292) corresponds to a frequency of 17 MHz. When the histograms are assembled as a function of time, we obtain the so-called dynamic spectrum shown in Fig. 20. We see that the ROLSES dynamic spectrum has prominent horizontal lines in the frequency range 16 – 22 MHz roughly corresponding to the histogram in Fig. 19. There are also some fainter lines up to 27 MHz. These horizontal lines represent ground transmitter signals from Earth that pass through the ionosphere and reach the ROLSES instrument just before landing on the Moon.

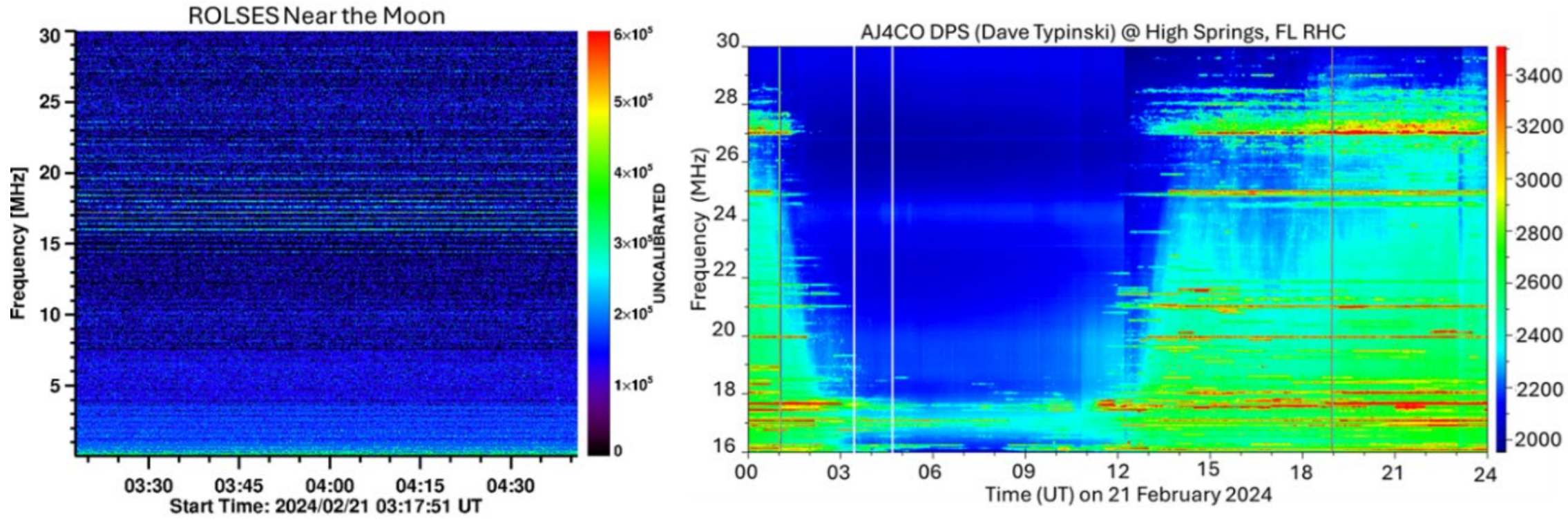

Fig. 20. (left) ROLSES dynamic spectrum (1-30 MHz) constructed using uncalibrated data taken during the transit when the lander was about 6000 miles from the Moon. There are bright horizontal lines (especially in the frequency range 15-20 MHz). (right) Dynamic spectrum (16-30 MHz) from the ground-based Radio Jove station AJ4CO located at Hot Springs, Florida obtained on February 21, 2024. The vertical white lines in the RJ dynamic spectrum mark the interval when ROLSES data were obtained. Note that it was nighttime in Florida during approximately 1-12 UT. North and Central America were facing towards the Moon at the time of the in-transit measurements (See Fig. 2 in Hibbard et al. 2026).

### *6.1 ROLSES and Radio Jove Dynamic Spectra*

As illustrated in Fig. 20 (left), Earth-based transmitters produced radio frequency interference (RFI) close to the Moon (or technosignatures that are useful for future SETI experiments). Note that there are not many horizontal lines below ~14 MHz. This is because the ionosphere blocks Earth-based radio signals lower than varying critical frequencies of the ionosphere. In Fig. 20 we compare the ROLSES radio observations with the simultaneous ground-based radio measurements in the frequency range of 16–32 MHz made by the AJ4CO observatory (https://aj4co.org/) located in High Springs, Florida, and operated by Dave Typinski. The AJ4CO Observatory is associated with the Radio JOVE (RJ) project (https://radiojove.gsfc.nasa.gov/). The AJ4CO spectrum was obtained by a dual polarization spectrograph. The RJ spectrum is for the whole day compared to just over an hour for ROLSES. We note that there were some radio transmitters detected by AJ4CO even during the night, mainly between 16-18 MHz and a couple of weak lines around 23 and 27 MHz. All these RFI lines are also discernible in the ROLSES dynamic spectrum.

For quantitative comparison, both ROLSES and AJ4CO/DPS data were further processed. The measured Radio Jove data contains non-conformities due to cable loss, spectrograph passband response, and antenna effects, and also includes the ever-present galactic background. These factors have been compensated by using a correction array that creates as uniformly flat frequency response as possible. The correction file contains one number per frequency channel per polarization that are added to the raw data values in the corresponding frequency channel. The correction array is created by measuring the observed signal level at each channel and polarization averaged over several minutes during quiet listening conditions, i.e., ideally observing only the galactic background. The low envelope of the observed signal is determined in order to remove possible other interfering signals. The low envelope of the measured signal is further smoothed by making a $5^{th}$ order polynomial fit to it. These fitted polynomial functions are then used to calculate the correction array (see as an example https://aj4co.org/AJ4CO Documentation Archive/Data Processing/AJ4CO DPS and TFD Correction Array Analsy 2017 01 10.xls). A correction array measured on 10 January 2017 was applied to the AJ4CO/DPS data used in this paper.

We divided the 80-minute measurement period from 03:20 UT to 04:40 UT into 4-minute intervals and obtained the time-integrated power spectrum in each interval for ROLSES cleaned data and AJ4CO/DPS corrected data. We detected the peaks using a sliding window method where the peak prominence was estimated by subtracting from the peak value the local background (the maximum of the two minimum values on either side of the peak within the time window). For both datasets, we adjusted the selection limit of the peak prominence to be reasonable, i.e., we excluded the tiny peaks. The limits are somewhat arbitrary, but we are interested in the stronger signals. Figure 21 compares the processed ROLSES dynamic spectrum with that from AJ4CO. We see clear horizontal lines above about 15 MHz in the ROLSES spectrum, similar to what was shown in Fig. 20. The AJ4CO corrected spectrum also shows a set of horizontal lines above 16 MHz. We matched each AJ4CO peak with the nearest ROLSES peak in each 4-minute interval. The counts of detected AJ4CO peaks within the 4-minute intervals at each AJ4CO frequency are plotted in Fig. 22. At lower frequencies, especially around 17 MHz, the frequencies of the two measurements match consistently, whereas at the higher frequencies above 19 MHz the matches are few and at the highest frequencies somewhat poor. The linear correlation between the two data sets is ~0.95, which significantly exceeds the Pearson critical correlation coefficient of 0.479 for 28 data points. The high correlation is an indication that the two instruments are observing the same transmitters. A more appropriate statistics may be the Kendall's rank correlation coefficient (also known as Kendall's Tau), which is a measure of association for two variables that can be ordered. Tau close to +1 indicates that the data pairs are mostly concordant. In our case, we have 161 data pairs, giving a correlation coefficient of 0.902, with p = 0.0000, indicating that the ROLSES and Radio Jove frequences show very similar ordering. Therefore, we believe that ROLSES was detecting radio signals originating from Earth during the lunar transfer.

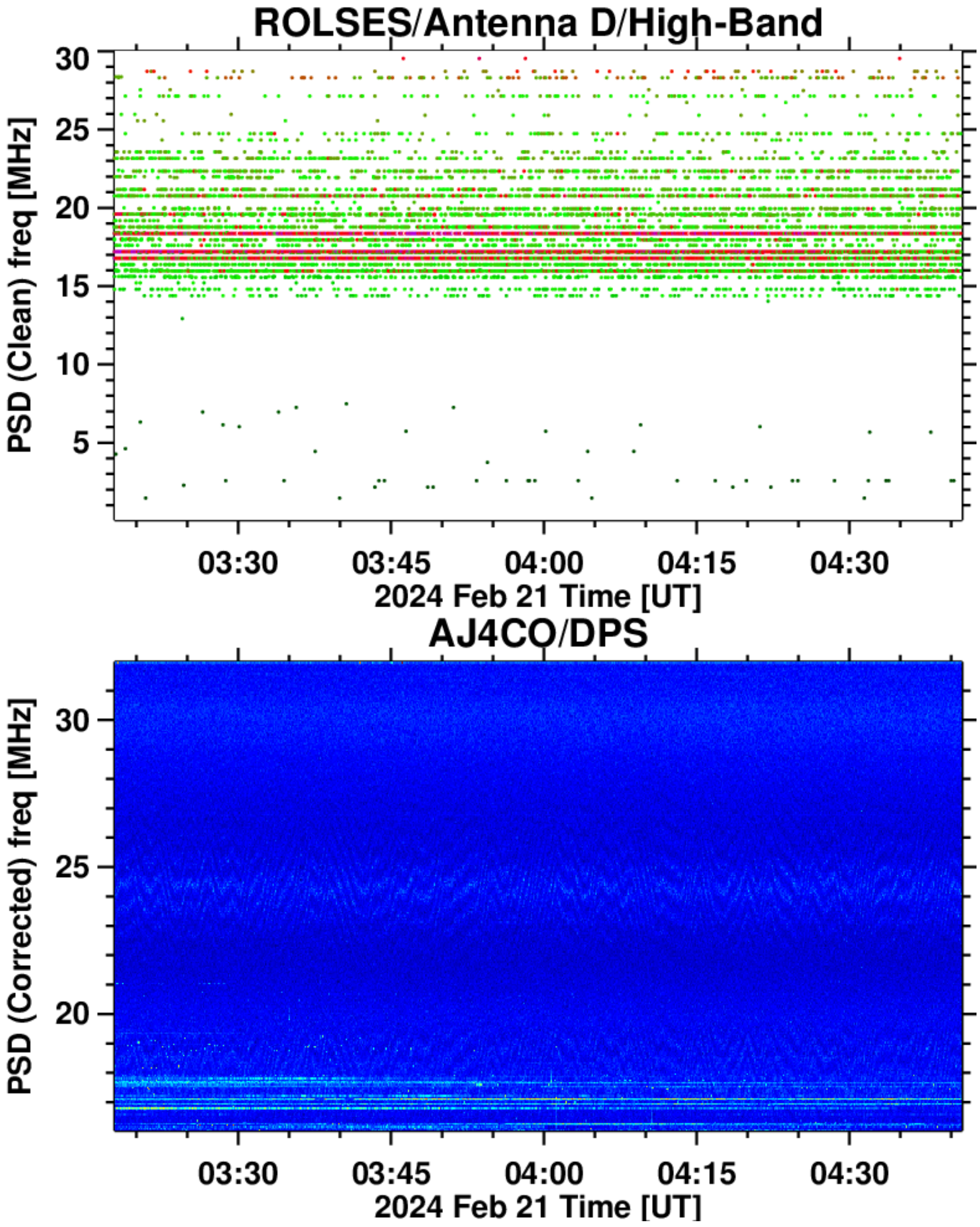

Fig. 21. Plots of power spectral density (PSD) obtained from ROLSES (upper panel) and the Radio Jove station AJ4CO (lower panel) on 2024 February 21. The ROLSES data are in the high frequency band from the Antenna that deployed in transit.

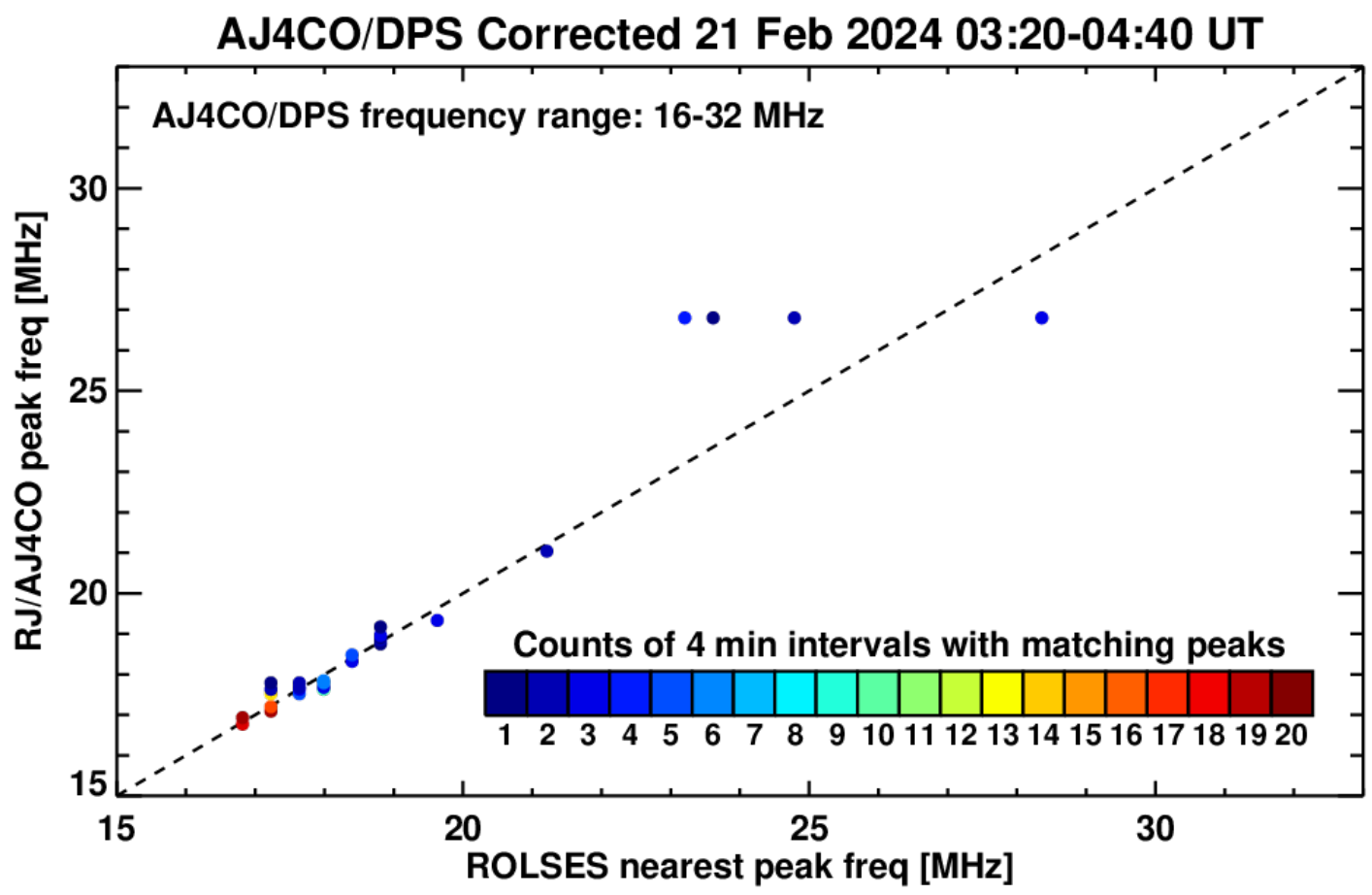

Fig. 22. Scatter plot between the ROLSES and Radio Jove (RJ) frequencies that show significant RFI. The data points correspond to counts of matching frequency peaks averaged over 4-min intervals in each spectra. The dashed line is the bisector. The color bar indicates the count values. The RJ spectrum is from the AJ4CO observatory in High Springs (29.8369° N, 82.6214° W) Florida run by Dave Typinski (amateur radio call sign AJ4CO) as part of NASA's NASA Radio Jove project (details: https://www.aj4co.org/AJ4CO%20Documentation%20Archive/Station%20Desriptions/AJ4CO%20Observatory%20-%20Station%20Description%202022%2001%2016.pdf).

### *6.2 Ionospheric Conditions Inferred from ROLSES Observations*

The Community Coordinated Modeling Center (CCMC) runs continuously the Coupled Thermosphere Ionosphere Plasmasphere Electrodynamics Model (CTIPe) and provides the simulation results available online at (https://iswa.ccmc.gsfc.nasa.gov/iswa_data_tree/model/ionosphere/CTIPe4.X/) through the Weather Analysis (ISWA) data tree. Fig. 23 shows the peak electron density of the F2 layer (NmF2) on 21 February 2024 at 04:00 UT. Such maps made every 10 min are available online at the above web site. The NmF2 parameter defines the minimum frequency ($foF2 = 9 \times 10^{-6} Nm^{1/2}$) for radio wave propagation from Earth to space. Here fo is in MHz and Nm is in electrons/m$^{-3}$. When a radio signal with frequency f is transmitted vertically, it will reflect back from the F2 layer when $f < fo$ or pass through the ionosphere when $f > fo$. For this reason, fo is called critical frequency. When the radio wave strikes the ionosphere at an angle, frequencies $> fo$ can be reflected and hence useful for long distance communication. The NmF2 map in Fig. 23 shows that the electron peak density above the central and eastern part of the U.S. ranged from $6.0\times10^{11}$ to $7.5\times10^{11}$ electrons/m$^{3}$, corresponding to the critical radio frequency (foF2) range from 7.0 to 7.8 MHz along the northern hemispheric longitudes that were facing towards spacecraft around 04:00 UT (see Fig. 2 in Hibbard et al., 2026). Depending on the inclination angle α between the vertical direction and the wave propagation direction, the maximum usable frequency (MUF, International Telecommunication Union 2015) scales as foF2/cosα. For example, if we assume that foF2=7.8 MHz then MUF will be ≈20 MHz when α≈53°. Therefore, radio waves in the frequency range of 16–19 MHz can either propagate into space or be reflected from the ionosphere depending on the inclination angle and the local ionospheric conditions.

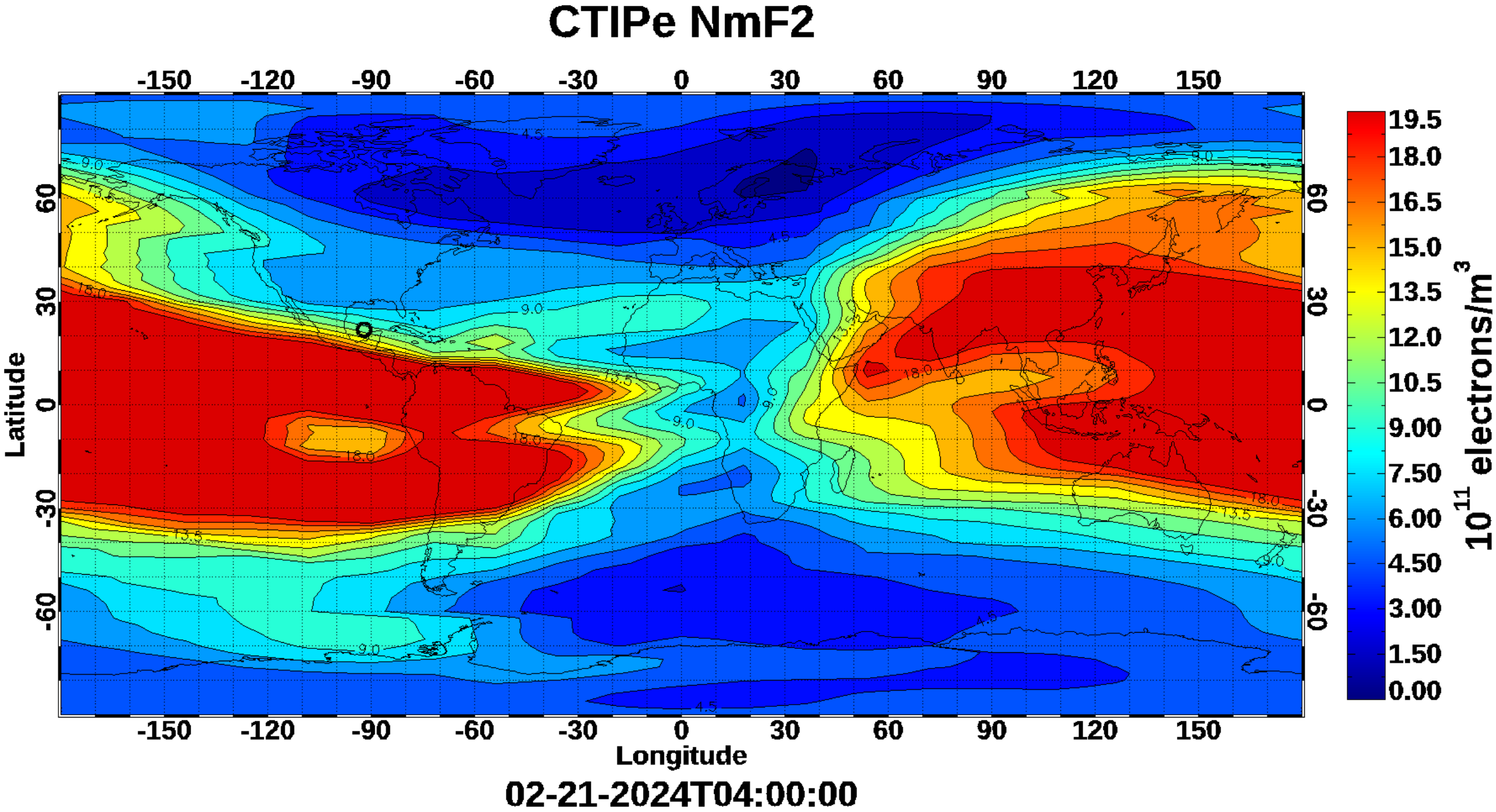

Fig. 23. Global distribution of the peak electron density of the F2 layer (NmF2) of Earth's ionosphere on 21 February 2024 at 04:00 UT. The projected location of the spacecraft is marked by an open circle near the Yucatan peninsula. The corresponding critical frequency foF2 of the ionosphere ranges from ~4.5 MHz to 18 MHz. Maps are available online at: https://iswa.ccmc.gsfc.nasa.gov/iswa_data_tree/model/ionosphere/CTIPe4.X/plots/NmF2

## 6.3 Lander Noise

Noise from the lander and the ROLSES electronics were quantified by measuring the output PSD of each antenna in various configurations: on the ground before launch when all antennas were stowed, in transit when one antenna was deployed, and on the lunar surface, before all antennas had deployed. If channels were found to contain high levels of electronic noise while deployed in any location, they were flagged as contaminated channels and removed from the analysis. The two dominant instrumental noise sources identified in the ROLSES-1 data were DSP quantization noise and front-end electronics (FEE)/lander noise. The DSP quantization noise arose from the bit-slicing architecture of the onboard digital signal processor: channels whose power failed to exceed the DSP trigger threshold produced only discrete quantized output values rather than true measurements. This effect was persistent, particularly in weak-signal regions of the spectrum, and constituted the majority of the raw DSP contamination before cleaning. In contrast, the FEE and lander noise originated from the antenna electronics, power systems, and electromagnetic interference generated by the lander itself. Ground-test measurements and observations taken while antennas remained stowed revealed stable narrowband contaminated channels that persisted across observing modes and epochs. They were thus identified in pre-flight and in-flight reference data and subsequently flagged and removed from the science observations. It must be noted that during pre-flight, the antennas could not be deployed for ground test as that would require refurbishment by the antenna manufacturer. The quantized noise was mostly concentrated in the low-band of the antennas (<5 MHz) and was at a known level. The remaining channels which exhibited noise during the ground test and before antenna deployment on the lunar surface were limited and only marginally above the quantized noise level, and again confined to the lower frequency ranges. The mid-band channels exhibited the least contamination from electronic and lander noise.

## 7. Future Work

A new and improved version of ROLSES (ROLSES 2) is being developed for launch in 2028. There are two primary additions to ROLSES 2: introduction of onboard calibration and modification of the DSP to obtain Stokes parameters. The primary purpose of the added calibration is to keep track of gain drifts with temperature and other environmental variables, using a programmable calibration signal generated inside the FPGA.

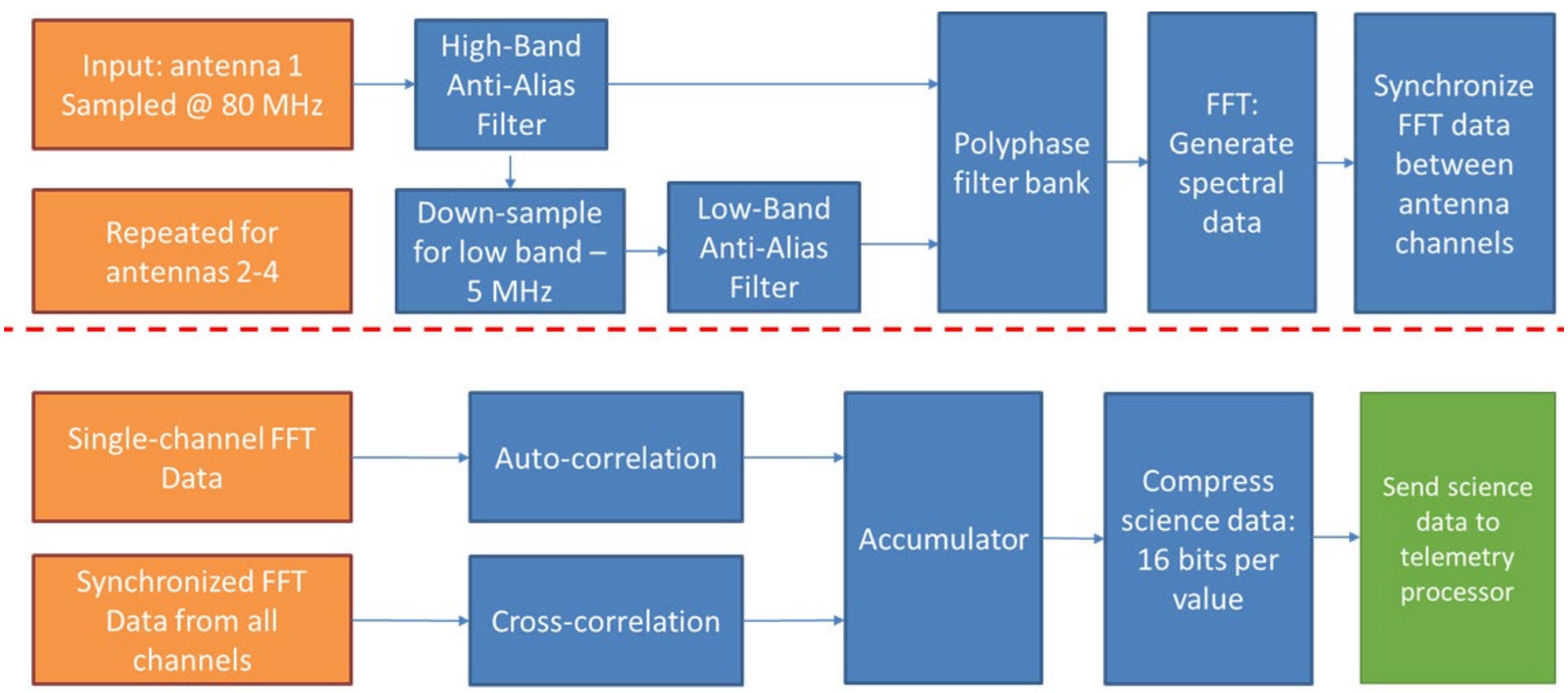

Fig. 24. A block diagram of the DSP chain for ROLSES 2. The top panel is nearly the same as in the previous version. The chains for the other antennas are identical. Both versions of ROLSES have four separate and identical FFT "cores." The controls allow us to enable or shut down each core individually (e.g., for faults, save power). Each of the four streams (antenna through science packets) is treated as an independent experiment in the upper portion.

The DSP algorithm for ROLSES 2 has been modified to (i) provide phase data between the antennas, (ii) fully meet FPGA timing requirements, and (iii) fix the data-discarding error from the flown version. The DSP chain for ROLSES 2 is shown in Fig. 24. The first portion (the blocks above the horizontal dashed line) is largely the same as in ROLSES 1: down-sampling each antenna's data stream as appropriate and generating spectral data with a polyphase filter bank and FFT processing. The sampling rate has been reduced from 120 MHz to 80 MHz to meet timing requirements within the FPGA. The high/low bands have been extended to 40 MHz and 2.5 MHz, respectively. At 80 MHz sampling, the length of 1024 samples is 12.8 microseconds, so it takes 839 milliseconds to perform 65536 FFTs in the high band and 4096 FFTs in the low band. The spectral resolutions are 78.125 kHz and 4.883 kHz, respectively for the high and low bands. All FFT data is then synchronized to a single digital clock domain to ensure accuracy of cross-correlation calculations required to obtain Stokes parameters on the ground. Each auto-correlation and cross-correlation is accumulated over up to 65536 iterations for the high band and up to 4096 iterations for the low band. The number of FFT iterations per second is configurable with ground commands. Rather than risk discarding any bits of data within the accumulator, the data's bit size is not reduced until the very end of the process. Finally, each data product is compressed to a 16-bit floating point value compliant with IEEE-754 half precision format giving high resolution at low levels while maintaining a large dynamic range. Moving to a 16-bit floating point format for ROLSES 2.0 when compared to the 32-bit format for ROLSES 1.0 enables doubling the rate of science measurements with the cost of some rounding for the larger signals, while maintaining high resolution for the lower level signals.

The placement of the antennas is also changed in ROLSES 2 to use the four monopoles as a crossed dipole. The position of each STACER will be co-linear with another STACER with no greater than a 5 cm parallel displacement and +/- 2 degrees linear offset. The second pair of co-linear STACER pair will be mounted orthogonal to the other co-linear pair with similar alignment limits. All the four STACERs will be in the same plane located at a height of 1-2 m from the lunar surface. Unlike ROLSES 1, test antennas very similar to flight antennas were procured for ROLSES 2.0 for pre-flight testing. These test antennas are in deployed state rather than a stowed/deployed flight antenna.
Finally, care has been taken to mitigate the inadvertent antenna deployment issues by using appropriate thermal blankets.

The communication architecture of ROLSES 2 will switch to the visualization tool Grafana, which is a multi-platform open-source analytics and interactive visualization web application. For ROLSES 2, it will produce charts, graphs, and alerts for the users. Database schema to store time series telemetry data and spectra has been demonstrated. Grafana instance with a live dashboard to see telemetry has also been demonstrated. A parser script has been developed to receive telemetry and spectral data and format the data to be stored in the database. Development is underway for Grafana custom plugins and scene for the commanding interface and other custom implementation.

ROLSES 2 will be one of the six payloads on Firefly Aerospace's Blue Ghost mission 3 (BGM 3) flying to Gruithuisen Gamma Dome located at 36°N, 40°W. BGM 3 is a lander-rover mission referred to as CLPS PRISM-2 call, 1st selection (CP-21, https://science.nasa.gov/lunar-science/clps-deliveries/cp-21-science/) scheduled for a 2028 launch. ROLSES 2 will be located on the Blue Ghost 3 lander.

## 8. Summary

ROLSES is a low frequency (< 30 MHz) radio telescope designed to characterize the complex radio radiation environment of the lunar surface for further exploration and development radio observatories on the Moon. ROLSES had a mass of 13.24 kg, average power of 22 W, and a data rate of 17 kbps. ROLSES was launched on board the IM-1 lander Nova-C (renamed Odysseus after landing). Upon landing the Odysseus lander tipped over and the power and telemetry for both downlink and uplink was very limited compared to that if it had landed as planned (https://nssdc.gsfc.nasa.gov/nmc/spacecraft/display.action?id=IM-1-NOVA). One of the antennas inadvertently deployed during transit; another inadvertently deployed on the surface. For each of these deployments, temperatures exceeded the specification limit due to unexpected excess solar flux enroute and on the Moon. The remaining two antennas were deployed by ground command; telemetry showed that these two deployments were nominal. Science data was taken in two forms: DSP processed waveforms for high and low bands, and raw data. As a result of limited commanding and operating times, the high gain settings could not be used during the mission. These data were sent to the lander, which downlinked the data for further processing on the ground. ROLSES observations were made when the Sun was very quiet, so no solar radio bursts were detected. This was verified by comparing ROLSES data with STEREO/WAVES dynamic spectra. Data taken during transit revealed the presence of terrestrial radio emissions from high frequency radio transmitters, as conformed by the Radio Jove dynamic spectrum temporally overlapping with ROLSES data. We confirmed that the ROLSES was also able to see the ionospheric cutoff. An upgraded ROLSES-2 is scheduled to go to the Moon in 2028 to complete the original science objectives for ROLSES.

**Acknowledgments**

We especially want to recognize the substantial contributions by co-author and former Principal Investigator Robert MacDowall who proposed ROLSES to the NPLP program and oversaw ROLSES instrument development through to its delivery. Shortly after delivery, his team members were devastated to learn of an acute and debilitating medical condition that forced his pre-mature retirement from the radio astronomy field he so deeply loved and which he contributed to over many decades. Dr. MacDowall passed away in March 2026. We, his team members, dedicate our efforts on ROLSES to our very dear colleague. This NPLP is supported by NASA's CLPS program and we thank them for their support throughout ROLSES development. We also thank Susan M. Lederer and J. Gruener (CLPS project scientists) and Francisco Moreno (CLPS integration Manager) for their support throughout this mission. We acknowledge the help and support from Maria Banks and Julie Schonfeld during ROLSES operations. Thanks are also due to Joanne Hill-Kittle and Beverly Settles who provided GSFC managerial support. We acknowledge the constant support provided by Joel Kearns (NASA HQ) and Ryan A. Stephan (NASA GRC) for the smooth progress of the ROLSES project. We thank Brad Costa and Greg Delory for answering questions related to the STACERs. We acknowledge data use from the Radio Jove station AJ4CO located at Hot Springs, Florida, and NASA CCMC. We thank Shing F. Fung for discussion related to the Radio Jove data. JB and JH acknowledge funding for data analysis of ROLSES from NASA grant 80NSSC23K0013.